\documentclass[11pt]{article}
\pdfoutput=1

\usepackage{amsmath, amsfonts, amssymb}
\usepackage{comment}
\usepackage{graphicx}
\usepackage{psfrag}
\usepackage[usenames,dvipsnames,svgnames,table]{xcolor}
\usepackage{enumerate}
\usepackage[utf8]{inputenc}
\usepackage{soul}
\usepackage{subfig}
\usepackage{mathrsfs}  
\usepackage{cite}
 \usepackage{a4wide}
   \usepackage{tikz}
  \usepackage{color}
  \definecolor{dark-gray}{gray}{0.20}
  \definecolor{gray}{gray}{0.30}
  \definecolor{light-gray}{gray}{0.80}
  \definecolor{dark-red}{rgb}{0.7,0,0}
  \definecolor{dark-green}{rgb}{0.1,0.4,0}
  \definecolor{dark-blue}{rgb}{0.3,0.3,0.7}
  \definecolor{light-blue}{rgb}{0.8,0.8,1}

\usepackage{setspace}

\newcommand{\be}{\begin{equation}}
\newcommand{\ee}{\end{equation}}

\def\be{\begin{equation}}
\def\ee{\end{equation}}
\def\bea{\begin{eqnarray}}
\def\eea{\end{eqnarray}}
\newcommand{\eq}[1]{(\ref{#1})}

\def\simleq{\; \raise0.3ex\hbox{$<$\kern-0.75em
      \raise-1.1ex\hbox{$\sim$}}\; }
   \def\simgeq{\; \raise0.3ex\hbox{$>$\kern-0.75em
      \raise-1.1ex\hbox{$\sim$}}\; }

\numberwithin{equation}{section}

\usepackage{hyperref}
\hypersetup{
	colorlinks=true,
	linkcolor=dark-blue,
	citecolor=dark-red,
	urlcolor=dark-green,
	linktoc=page
}
\begin{document}

\begin{center}

{\LARGE {\bf  Festina Lente\footnote{Latin expression which translates to ``Hasten slowly''.}: EFT Constraints from Charged Black Hole Evaporation in de Sitter}} \\
\vspace{0.5cm}

\vspace{1.5 cm} {\large  Miguel Montero,$^{a,b}$  Thomas Van Riet$^{a}$ and Victoria Venken$^{a}$ }\\
\vspace{0.5 cm}  \vspace{.15 cm} {$^{a}$Institute of Theoretical Physics, KU Leuven,\\
Celestijnenlaan 200D B-3001 Leuven, Belgium
}

\vspace{0.5 cm}  \vspace{.15 cm} {$^{b}$Jefferson Physical Laboratory, Harvard University\\
Cambridge, MA 02138, USA}

\vspace{0.7cm} {\small \upshape\ttfamily  mmontero @ g.harvard.edu,  thomas.vanriet @ kuleuven.be, victoria.venken@gmail.com
 }  \\

\vspace{2cm}

{\bf Abstract}
\end{center}
{ In the Swampland philosophy of constraining EFTs from black hole mechanics we study charged black hole evaporation in de Sitter space. We establish how the black hole mass and charge change over time due to both Hawking radiation and Schwinger pair production as a function of the masses and charges of the elementary particles in the theory. We find a lower bound on the mass of charged particles by demanding that large charged black holes evaporate back to empty de Sitter space, in accordance with the thermal picture of the de Sitter static patch. This bound is satisfied by the charged spectrum of the Standard Model. We discuss phenomenological implications for the cosmological hierarchy problem and inflation. Enforcing the thermal picture also leads to a heuristic remnant argument for the Weak Gravity Conjecture in de Sitter space, where the usual kinematic arguments do not work. We also comment on a possible relation between WGC and universal bounds on equilibration times. All in all, charged black holes in de Sitter should make haste to evaporate, but they should not rush it\footnote{Hence the title.}.}

\setcounter{tocdepth}{2}
\newpage

\tableofcontents

\section{Introduction}

The Swampland program \cite{Ooguri:2006in,nLab} (see \cite{Brennan:2017rbf,Palti:2019pca}) aims to discover, itemize and establish the set of consistency conditions imposed on a low-energy Effective Field Theory (EFT) by demanding that it can consistently coupled to quantum gravity. The current status of the program is a plethora of proposed Swampland constraints with varying degrees of support, ranging from absence of global symmetries \cite{Banks:1988yz,Abbott:1989jw,Coleman:1989zu,Kallosh:1995hi,Banks:2010zn,Susskind:1995da,Beem:2014zpa,Harlow:2018jwu,Harlow:2018tng} (see \cite{Fichet:2019ugl} for proposals on how strong should the breaking be), the Weak Gravity Conjecture and all of its variants \cite{ArkaniHamed:2006dz,Cheung:2014vva,Heidenreich:2016aqi,Andriolo:2018lvp,Klaewer:2016kiy,Palti:2017elp,Lust:2017wrl,Montero:2016tif,Reece:2018zvv,Lee:2018urn,Lee:2018spm,Bonnefoy:2018mqb,Klaewer:2018yxi,Hamada:2018dde,deRham:2018dqm,Hebecker:2018yxs,Craig:2018yvw,Bellazzini:2019xts,Pal:2019pqh,Loges:2019jzs,Cheung:2019cwi,Brahma:2019mdd,Craig:2019fdy,Kaplan:2019soo,Shirai:2019tgr,Mirbabayi:2019iae,Nath:2019yna,Charles:2019qqt,Gonzalo:2019gjp,Lee:2019tst,Lee:2019xtm,Aalsma:2019ryi,Heidenreich:2019zkl,Ong:2019glf,deAlwis:2019aud,Remmen:2019cyz,Jones:2019nev,Goon:2019faz,Cano:2019oma,Ong:2019rnn}, or Swampland Distance Conjecture \cite{Ooguri:2006in,Heidenreich:2017sim,Grimm:2018ohb,Heidenreich:2018kpg,Grimm:2018cpv,Lin:2018edm,Buratti:2018xjt,Hebecker:2018fln,Gonzalo:2018guu,Scalisi:2018eaz,Marchesano:2019ifh,Kehagias:2019akr,Corvilain:2018lgw,Font:2019cxq,Grimm:2019wtx,Gautason:2018gln,Lust:2019zwm}, to the conjecture that there are no weakly coupled long-lived de Sitter solutions \cite{Obied:2018sgi,Agrawal:2018own,Kinney:2018kew,Kiritsis:2019wyk,Goswami:2018pvk,Kaloper:2019lpl,Gomez:2019ltc,Bedroya:2019snp}. Some of these conjectures are supported by heuristic black hole arguments \cite{Susskind:1995da} or independent holographic evidence (in the AdS context) \cite{Harlow:2018jwu,Harlow:2018tng,Montero:2018fns}, but all of them are ultimately relying on a vast amount of examples from string compactifications (or, in the dS or non-supersymmetric AdS cases, lack thereof \cite{Sethi:2017phn,Danielsson:2018ztv,Obied:2018sgi}). 

The ample evidence for some Swampland constraints in string theory gives us confidence they are probably correct, but it is equally important that at least in some cases we seem to have identified a physical principle underlying the conjecture. For instance, the conjecture that there are no global symmetries, which has been shown to hold in detail in AdS/CFT \cite{Harlow:2018jwu,Harlow:2018tng}, was originally motivated by UV-insensitive arguments about black holes and remnants. The fact that the conjecture seems to hold in every example could be taken as evidence that the original heuristics was probably correct. If the principle is correct, we can consider applying it in situations where it is not firmly established. For instance, consider $(B-L)$. This is a global symmetry of the SM lagrangian, and based on the above one would imagine that is gauged or broken in the real world,  even if nobody has been able to get the SM or even positive vacuum energy in string theory or other would-be UV-complete framework. This is because we do have black holes and the original argument \cite{Banks:1988yz,Susskind:1995da,Banks:2012hx} which relies on getting stable remnants by throwing baryons into black holes seems to apply. 

A similar story applies to the Weak Gravity Conjecture. In a very handwavy way, the principle underlying WGC is that something bad happens if large extremal black holes cannot decay into slightly smaller (sub-)extremal black holes\footnote{In the supersymmetric case, this can happen only marginally, and whether the decay actually takes place or not depends on the detailed dynamics (e.g. walls of marginal stability). What seems to matter is that the decay is kinematically allowed (a large BPS object can fragment into smaller ones). }. So the principle seems to be something like demanding that black holes should decay while remaining sub-extremal; this connection is made more precise in \cite{Crisford:2017zpi,Crisford:2017gsb,Horowitz:2019eum}, which show that adding WGC particles to a particular theory in AdS is enough to comply with another physical principle, Weak Cosmic Censorship.

If we got the principle right, again we can hope to apply it in situations where we cannot validate it independently. In this paper we take the first steps towards generalizing WGC-like arguments to spacetimes with positive vacuum energy. Our arguments will take place in de Sitter, since this is the simplest example, but we imagine (and have checked in some cases) that they hold more generally, like in slow-roll quintessence models. 

The first step is to understand black hole evaporation in de Sitter. This is more complicated than in flat space since there is both a black hole and a cosmological horizon, and they exchange charge and mass. The problem was worked out early on for neutral black holes  \cite{Bousso:1996au,Bousso:1996pn,Bousso:1997wi,Bousso:1998bn} and charged ones in the absence of charged particles \cite{Bousso:1999ms,Bousso:2002fq}. While there are some works that discuss emission of charged particles by black holes in de Sitter \cite{Belgiorno:2008mx,Belgiorno:2009da,Belgiorno:2009pq,Kim:2015wda,Belgiorno:2016nup}, we were unable to find a reference that described in detail how the mass and charge of the black hole deplete. We do this in the present paper, following an approach similar to \cite{Hiscock:1990ex}, although the details are more complicated due to the lack of an asymptotically flat region. 

What we will find is that, as long as the charged carrier is heavy enough  in the sense that, 
\begin{equation*} m^2 \gg qg  M_P H,\end{equation*}
(but still satisfying the WGC inequality) black holes in dS always evaporate all the way to empty de Sitter space. Here, $m,q$ are the mass and charge of the elementary particle that is discharging the black hole, $g$ is the $U(1)$ gauge coupling, $M_P$ is Planck's mass and $H$ is the Hubble scale.  This does not come as a surprise \cite{Dias:2018etb}, since unlike in flat space a charged particle inside the black hole can always tunnel close to or behind the cosmological horizon, therefore discharging the black hole. This result fits right in with another principle that seems to hold in de Sitter, namely that physics in the static patch behaves very much like a finite-dimensional thermal system at an equilibrium temperature of $H/(2\pi)$ \cite{Anninos:2012qw}. Any state different from the vacuum in the static patch corresponds to taking the system out of equilibrium, and the system responds by eventually re-equilibrating. This is the case for small perturbations, small black holes, and even neutral and charged black holes in the absence of charged particles\footnote{We note in passing the recent interest on the Schottky anomaly \cite{Dinsmore:2019elr,Johnson:2019ayc}, a feature typically seen in the heat capacity of finite dimensional thermal systems which has an avatar in black holes in de Sitter space as well. This provides further evidence for the consistency of the thermal picture in dS.} \cite{Bousso:2002fq}. We establish that this holds in the presence of heavy enough charged particles as well. This bound does not become trivial in the $M_P \rightarrow \infty$ limit. We will comment further on this in the conclusions. 

However, something funny happens in the opposite limit of a very light charged particle (in the sense that $m^2\ll qg M_P H$).  In this case, we find that very large charged black holes (so large that the black hole and cosmological horizons coincide; these are called Nariai black holes \cite{Nariai1999OnSS,Romans:1991nq}) discharge essentially instantaneously and, due to the fact that  charged Nariai solutions are more massive than neutral Nariai solutions (see Figure \ref{fig1}), they collapse into a Big Crunch, rather than returning to empty de Sitter space. Due to the quick discharge, the discharging charged Nariai solutions resemble ``superextremal'' Nariai solutions.

Thus, if there is at least one light charged particle in the theory, the thermodynamic picture of de Sitter in the static patch cannot hold, large charged black holes never re-equilibrate, and one gets outside of the sub-extremal region of allowed black holes in Figure \ref{fig1}. This might be a bad thing and so we entertain the possibility that $m^2\gtrsim qg M_P H$ might be an actual constraint on the EFT. We emphasize that due to the lack of stringy examples this extrapolation is in shakier grounds than other constraints such as WGC or absence of global symmetries, since we cannot check it; all we can say is that the principles we extrapolate (thermality of the dS static patch, no superextremal black hole solutions) seem a reasonable extrapolation of what quantum mechanics and gravity in de Sitter might look like. Nonetheless the findings and computations carried out in this paper are interesting from a GR and QFT viewpoint, regardless of whether these principles survive eternally or not.

It turns out that avoiding fast discharge of Nariai solutions also leads to WGC. There are two ways to trigger a quick black hole discharge, either by having a few light particles in the above sense, or by having a huge number of heavy ones, so that the black hole discharges via their combined effect. In a theory in which the usual WGC is not satisfied, small black holes are very long-lived and play exactly this role in the limit of small weak coupling. Thus, one recovers a form of the original WGC heuristics that works in dS space.

A constraint like $m^2\gtrsim qg M_P H$ can be checked against the real world, since it looks like de Sitter. Taking the $U(1)$ to be electromagnetism, we find it is satisfied by every charged particle in the Standard Model. One could take this as evidence for the thermal picture of the dS static patch. Since in the SM the fermions get their masses from coupling to the Higgs, the inequality alleviates (but does not solve) the cosmological hierarchy problem. It leads to no new constraints on models of milli-charged dark matter. There is an interesting interplay with inflation, where the constraint is satisfied  by only some inflationary models. Possibilities include small field inflation, direct gauge kinetic term-inflaton coupling, or (tuned) models of Higgs inflation.

The paper is organized as follows: 
\begin{itemize}
\item Section \ref{sec:pre} reviews the charged black hole solutions of the Einstein-Maxwell-deSitter theory and the relevant near-horizon limits.
\item Section \ref{sec:decay} is the core of the paper. We analyze black hole discharge via Schwinger pair production of charged particles, both in the adiabatic and quasistatic regimes. 
\item Section \ref{sec:conjs} discusses the failure of thermality/detailed balance in the static patch in the adiabatic discharge regime. We suggest that this is evidence that the adiabatic regime could be pathological and should be avoided, discussing the phenomenological implications. We also explain how this is connected with WGC and discuss how the connection could be sharpened.
\item Section \ref{sec:conclus} present our conclusions and afterthoughts. 
\end{itemize}
Finally, some Appendices contain details of the calculations and some elaborations on the arguments we present.

\section{Charged black holes in de Sitter space}\label{sec:pre}
In this Section we review the black hole solutions we will be interested in. We consider a $(3+1)$-dimensional Einstein-Maxwell-de Sitter system, with action (in $-+++$ signature)
\begin{equation}S=\int d^4x\sqrt{-g}\left[ \frac{1}{16\pi G}\left(-R+\frac{6}{\ell^2}\right)+ \frac{1}{4g^2}F_{\mu\nu}F^{\mu\nu} \right].\end{equation}
This theory admits charged black hole solutions, the RN-dS metric, (see e.g. \cite{Romans:1991nq}):
\begin{equation}ds^2=-U(r) dt^2+\frac{dr^2}{U(r)} +r^2 d\Omega,\label{RNdSmetric}\end{equation}
with
\begin{equation}U(r)\equiv 1-\frac{2GM_r}{r} + \frac{G(gQ_r)^2}{4\pi r^2}-\frac{r^2}{\ell^2}.\label{Ur0}\end{equation} 
This metric is supported by an electric field or magnetic field. In the electric case we have 
\begin{equation}A=\Phi\, dt, \quad \Phi=\frac{g^2}{4\pi} \frac{Q_r}{r}.\label{A}\end{equation}
The two parameters $M_r$ and $Q_r$ can be interpreted as a ``mass'' (more about this in Section \ref{sec:decay}) and the charge. In de Sitter space the notion of mass is ambiguous but, language wise we will refrain from these subtleties and will name $M_r$ the mass from here onwards.  In this normalization, a particle of charge $q_r$ couples to the electrostatic potential via
\begin{equation} q_r \int A.\end{equation}
If we choose $g$ such that the particle of lowest charge in the spectrum has $q_r=1$, then charges are integer-quantized.

From now on we will work in Hubble units, $\ell=1$. Then the metric above takes the form
\begin{equation}U(r)\equiv 1-\frac{2M}{r} + \frac{Q^2}{r^2}-r^2,\label{Ur}\end{equation} 
with the dimensionless parameters
\begin{equation} M\equiv \frac{GM_r}{\ell},\quad Q^2\equiv \frac{Gg^2Q_r^2}{4\pi \ell^2}.\label{adimp}\end{equation}
\begin{figure}[!htb]
\begin{center}
\includegraphics[width=0.55\textwidth]{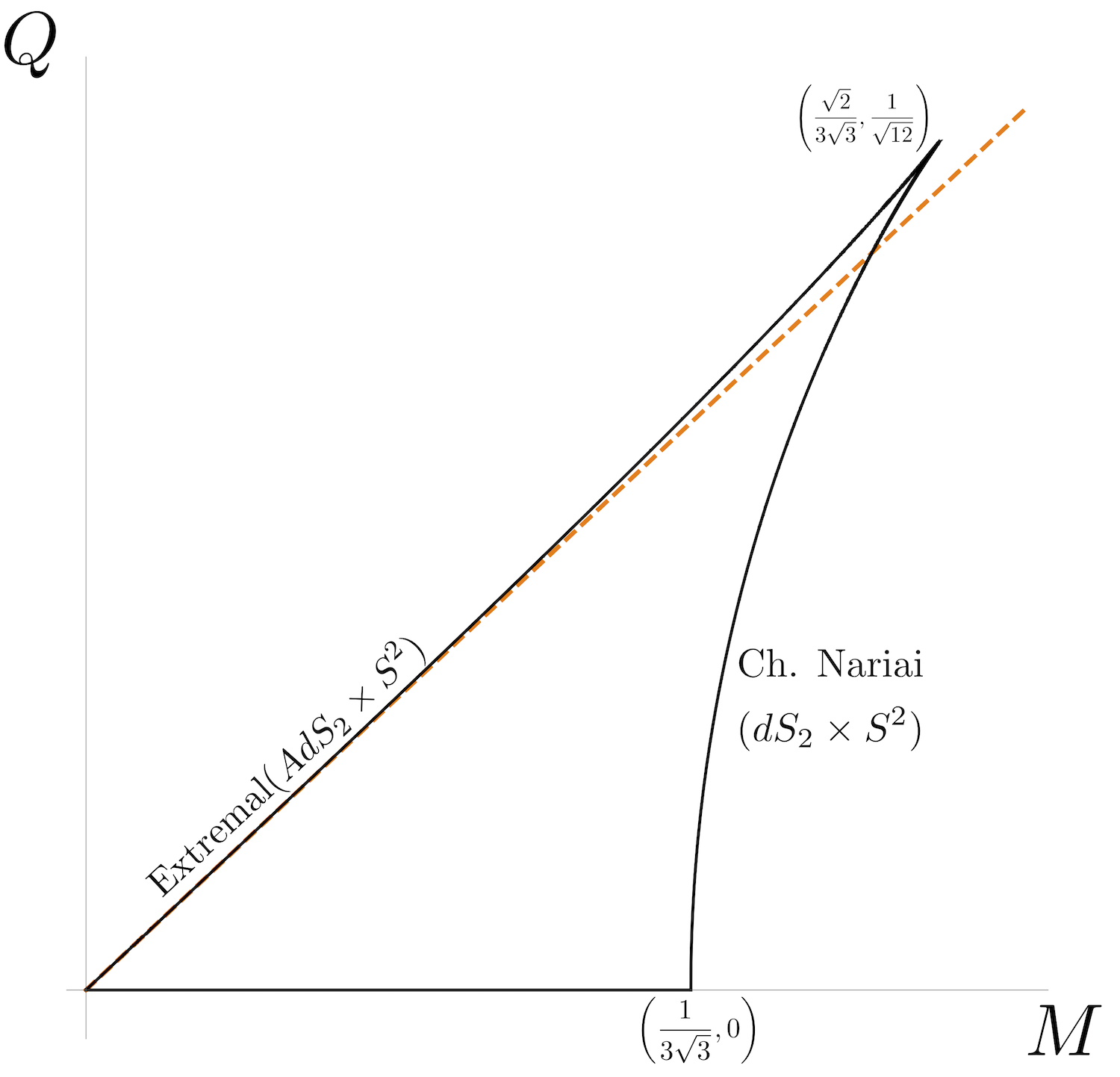}
\vspace{0.3cm}
\caption{\small{Phase diagram of Reissner-Nordstrom-de Sitter black holes, the ``shark fin''. Outside of the curve of extremal solutions, the metric becomes superextremal and has a naked singularity. The boundary has two branches: Extremal RN-dS black holes with an $AdS_2\times S^2$ horizon topology and charged Nariai black holes for which the black hole and cosmological horizons coincide, with a $dS_2\times S^2$ near-horizon geometry. The orange dashed line is the ``lukewarm line'' $Q=M$, where the temperatures of the black hole and cosmological horizons are identical.}}\label{fig1}
\end{center}
\end{figure}
The RN-dS metric \eq{Ur0} contains two horizons accessible to an observer outside the black hole\footnote{In addition to these, the black hole has an inner horizon $r_-$ when subextremally charged.}: one is the usual event horizon of the charged black hole, and the other is the cosmological horizon. In general, the black hole and the cosmological horizons will have different temperatures, and so they won't be in thermal equilibrium. When the black hole is large enough, the system will slowly drift towards equilibrium. This will  always increase the entropy, whose leading contribution comes from the two horizons,
\begin{equation} S= \frac{\pi}{4G}\left( r_{BH}^2+r_{CH}^2\right),\end{equation}
but how this happens exactly depends on the particular details of the dynamics of the system. The solution \eq{Ur0} has two parameters $M$ and $Q$, and the phase diagram as a function of $M$ and $Q$ (which we will call the ``shark fin'') is depicted in Figure \ref{fig1}.  There is a boundary defined by the discriminant locus $\Delta=0$ of the quartic equation $U(r)=0$,where
\begin{equation}\Delta \equiv M^2- Q^2-27 M^4+36 M^2 Q^2-8  Q^4-16 Q^6.\label{discr}\end{equation}
 This boundary comprises extremal black hole solutions, which in turn split into two different branches \cite{Romans:1991nq}:
\begin{itemize}
\item The upper branch of the blue curve in Figure \ref{fig1} is called the extremal branch, and parametrizes extremal black hole at zero temperature with horizons smaller than the cosmic horizon. The inner and outer black hole horizon coincide, $r_+=r_-$. The near-horizon geometry is $AdS_2\times S^2$. 
\item The lower branch of the curve in Figure \ref{fig1} is the charged Nariai branch, and it contains subextremal charged black holes with the same area as the cosmological horizon. The outer black hole horizon and cosmic horizon are in thermal equilibrium and appear to coincide, $r_+=r_c$, though a coordinate reparametrization will show this not to be the case. The near-horizon geometry is $dS_2\times S^2$ and the temperature increases as one moves down the branch. We will be especially interested in this branch and discuss it in greater detail below.
\end{itemize}
These two branches meet at a point, dubbed the ``ultracold'' black hole \cite{Romans:1991nq} since semiclassically it describes an equilibrium configuration at vanishingly small temperature. Here, $r_-=r_+=r_c$. The near horizon geometry is $\mathbb{M}_2\times S^2$, exactly interpolating between the two branches. Outside of the extremal $\Delta=0$ curve in the diagram one always has naked singularities, except on the $Q=0$ axis of neutral black holes, where one has a Big Crunch singularity instead. 

We want to study black holes for any values of $M,Q$ in the shark fin. It would be convenient to have some coordinate-independent description which works smoothly even near the boundaries of the diagram. An interesting possibility is what we will call the ``geodesic observer'', sitting at a particular value of the radius $r_g$. This is the radius between the black hole and the cosmic horizon such that the pull of the black hole and cosmic expansion cancel, $U'(r_g)=0$. An observer at $r=r_g$ can follow an orbit of the timelike Killing vector field which is at the same time a geodesic. Stuff closer to the black hole than the geodesic observer will eventually fall in, while stuff further out will keep on receding from the observer towards the cosmological horizon. 

\subsection{The charged Nariai branch}
As black hole mass increases and a black hole approaches the Nariai branch, $r_+ \rightarrow r_c$. This is a near horizon limit, in which $U(r)$ develops a double zero, so as usual to properly understand this Nariai limit, we need a change of coordinate. Let
\begin{equation} \rho= \frac{r-r_g}{\sqrt{\vert U(r_g)\vert }},\quad \tau= \sqrt{\vert U(r_g)\vert } t.\label{hcs}\end{equation}
The RNdS metric in terms of these radial and time coordinates is given by
\begin{equation} ds^2=-\frac{U(r)}{\vert U(r_g)\vert } d\tau^2+ \frac{\vert U(r_g)\vert }{U(r)} d\rho^2+ r^2d\Omega^2, \label{metric-nariai}\end{equation}
The constant electric field has the same magnitude in these coordinates,
\begin{equation} F= F_0\, dr\wedge dt = F_0\, d\rho \wedge d\tau,\quad F_0=\frac{g^2}{4\pi} \frac{Q_r}{r^2}.\end{equation}

 On this coordinate chart, the transition from subextremal to extremal black is completely smooth, and the horizons never overlap. Even though in the coordinates \eq{RNdSmetric} the Nariai branch corresponds to the limit $r_+=r_c$, we see that from the point of view of the geodesic observer the two horizons do not actually collide, although they do get closer as we approach the charged Nariai branch. Exactly on this branch, the metric components in \eq{metric-nariai} simplify to
\begin{equation}\frac{U(r)}{U(r_g)} \rightarrow 1-\frac{\rho^2}{\ell_{dS_2}^2},\quad r^2\rightarrow r_c^2,\label{ds2met}\end{equation}
which (as is well-known \cite{Anninos:2012qw}) is exactly the $dS_2\times S^2$ metric with a $dS_2$ radius
\begin{equation}\ell_{dS_2}^2=\frac{2}{U''(r_c)}=\frac{1}{6} \left(\frac{1}{\sqrt{1-12 Q^2}}+1\right)=\left(3-\frac{Q^2}{r_c^4}\right)^{-1}.\label{hubble2d}\end{equation}
The radius of the $S^2$ on the charged Nariai branch is given by
\begin{equation} r_c(Q)=\sqrt{\frac16\left(1+\sqrt{1-12 Q^2}\right)}.\label{rc2d}\end{equation}
Physically, as we approach extremality, the usual behavior of near-horizon limits sets up, where the geometry is approximately $dS_2\times S^2$. The geodesic observer happens to fall right in the middle of this tube, and the coordinates \eq{hcs} are just local coordinates adapted to her. 
\section{Semiclassical evolution of black hole solutions}\label{sec:decay}

Classically, the black hole solutions we have just discussed are stable\footnote{Classical stability is controlled by the spectrum of quasi-normal modes on the classical solution. These have been the subject of recent study \cite{Dias:2018ufh} in connection with variants of Strong Cosmic Censorship. The takeaway message for us is that for 4-dimensional black holes there are no instabilities, except for scalar fields of very low mass and nonvanishing charge. In this case, the instabilities only appear in a region of the $(M,Q)$ plane far from the extremality curves, so it is not of direct relevance to us.}, but quantum mechanically, there is backreaction on both the metric and gauge field. This is because there is Schwinger and Hawking radiation coming out of each horizon. 

 Let us ignore for a moment Schwinger radiation and focus on the dynamics as dictated solely by Hawking radiation. Since for generic values of $M$ and $Q$ the two horizons have different temperature, we expect the system to slowly drift towards equilibrium; in other words, the ``mass'' $M$ will evolve towards its equilibrium value slowly \cite{Bousso:1997wi}. Just like in flat space, the backreaction of Hawking radiation is only significant for a Planckian black hole.  

Unlike in flat space, however, there is Hawking radiation coming out of both horizons, and generically, it is at different temperatures. The dashed orange line $M=Q$ in Figure \ref{fig1} describes the line of ``lukewarm'' black holes \cite{Romans:1991nq}, where the Hawking radiation coming from the black hole and cosmological horizons are at the same temperature. If we neglect charged particles, any point on this line is at thermodynamic equilibrium; we expect black holes to slowly evolve along lines of constant charge towards the lukewarm solution. 

When one includes the effects of charged particles, which feel the electric field stretching between the two horizons, even this line becomes unstable. One can understand this by describing the horizons as reservoirs in the grand canonical ensemble \cite{Gibbons:1975kk}; to reach equilibrium, both the temperatures and electrostatic potentials of the two horizons must be identical. On the lukewarm line, this only happens at the lower left corner, that is, empty $dS$ space. 

The physical process by which a black hole loses charge is Schwinger radiation \cite{Schwinger:1951nm,Gibbons:1975kk,Hiscock:1990ex}. The black hole spacetime contains an electric field, and at any point, charged carriers are pair produced from the vacuum at some rate dependent on the local strength of the electric field. In flat space, this transition rate has an exponential suppression,
\begin{equation}\Gamma\sim e^{-\frac{m^2}{qE}},\label{schw00}\end{equation}
controlled by the mass of the charged particle and the electric field. Our task is to understand how the black hole charge, $Q$, changes as a function of the Schwinger current. Unlike Hawking radiation, whose energy flux represents a small fraction of the black hole mass for any but the smallest black holes, the Schwinger current can be huge or tiny compared to the background electric field, and this is controlled roughly speaking by \eq{schw00}. Thus, there are two regimes, depending on whether $m^2\lessgtr qE$, which we will discuss separately.

\subsection{Quasistatic discharge: \texorpdfstring{$m^2\gg qE$}{m2>qE}}
We will start with the regime in which the pair production \eq{schw00} is very much suppressed. In this regime, Schwinger radiation behaves just like Hawking -- the outgoing charge flux is small compared to the background charge. A similar analysis was carried out in \cite{Hiscock:1990ex} for charged black holes in flat space. We will reproduce their equations of motion in the appropriate limit. 

A significant difference between the flat space and the dS cases is that in flat space one has conserved charges at infinity (total charge and mass). The problem is therefore just a matter of computing these fluxes at infinity. In de Sitter, on the other hand, there is no spatial asymptotia, and we must instead work with Einstein's equations directly. We will show that, under quasistatic evolution, the metric is very well approximated by \eq{RNdSmetric} and \eq{Ur0} with slowly-varying $(M,Q)$ as a function of time, just like in the flat space case. 

Let us describe the slow drift of the solutions on the $(M,Q)$ plane in more detail. This can be traced to the usual nonzero vevs $\langle T_{ab}\rangle$ and $\langle j_a\rangle$ for quantum field theory in curved spacetimes. We want to understand the dynamics of this system in the quasi-static regime. To do so, we will solve the equations of motion perturbatively,
\begin{equation} \delta G_{ab}=8\pi G\, \delta T_{ab},\quad d\star \delta F=\star\langle \delta j\rangle,\label{eep}\end{equation}
to first order in the perturbations 
\begin{equation}
\delta T_{ab}=\langle T_{ab}\rangle - T_{ab}^{\text{classical}}\,, \qquad \delta j= \langle j \rangle - j^{\text{classical}}\,.
\end{equation}
For the RN-dS solutions, $ j^{\text{classical}}=0$, while $T_{ab}^{\text{classical}}$ is due to the background electromagnetic field.To ease bookkeeping, we will introduce a small dimensionless parameter $\epsilon$
\begin{equation}\epsilon\equiv G \ell^2 \sqrt{\delta T_{a}^b\delta T_b^a}\end{equation}
of the order of the largest matrix element in $\delta T_{ab}$ (in the coordinate system \eqref{Ur0}). So for instance, for thermal radiation, $\epsilon\sim T^4/(M_P H)^2$. We will also assume that the current is of order $\epsilon$ or smaller. 

Then, we will solve \eq{eep} in a quasi-static approximation, using the method of separation of scales (see e.g. \cite{kevorkian1996multiple}); since there is a whole two-parameter family of solutions \eq{Ur0} to the unperturbed equations of motion, the perturbation will lead to slow motion in this parameter space, similarly to the computation of the perihilion precession in GR. More technically, one could work in the $\epsilon$ expansion of \eq{eep}. Since the metric is $t$-independent, it is possible to separate variables,
\begin{equation}\delta g \sim e^{i\omega t} g(r),\end{equation}
and then solve the linearized problems order by order to any desired accuracy in $\epsilon$. If the spectrum of $\omega$'s is gapped, then the perturbation becomes a Fourier transform and under reasonable conditions it remains bounded (order by order in $\epsilon$) at all times. However, if there are zero modes (secular perturbations), then they will grow linearly (as $\epsilon^n t$ at order $n$), spoiling agreement between the unperturbed and perturbed problems at a time $\epsilon^{-n}$. The mass and charge parameters in \eq{Ur0} are examples of these zero modes. 

The method of separation of scales removes the secular dependence of zero modes and works as follows. We introduce a ``slow scale'' 
\begin{equation} t_1\equiv \epsilon\, t\end{equation}
on which the parameters of the background metric may depend,
\begin{align}M&\rightarrow M(t_1),\nonumber\\Q&\rightarrow Q(t_1).\label{perturanst}\end{align}
We also introduce an order $\epsilon$ perturbation\footnote{This is not the most general perturbation possible, but it will be enough to solve the equations of motion.},
\begin{equation} \epsilon^{-1}\delta ds^2= \delta A(r,t_1) dr^2+ \delta B(r,t_1) dt^2, \label{perturanst2}\end{equation} 
and a similar perturbation to the electric field,
\begin{equation} F\rightarrow F+\epsilon g\ell\sqrt{G} \delta F(r,t_1) dr\wedge dt.\end{equation}
 A perturbation like \eq{perturanst} and \eq{perturanst2} on the metric results in a first-order perturbation to the Einstein tensor $\delta G_{ab}$ which is covariantly conserved with respect to the background metric, is time independent, and respects the symmetries of the background. The most general such tensor is of the form
\begin{equation} \delta G_{ab}- T_{ab}^{(0)}=\delta_{1}\, \xi_a \xi_b+  \delta_{2}\,v_av_b + \delta_3\, \xi_{(a}v_{b)} + \delta_4\, g_{ab},\label{expandt}\end{equation}
 where $\xi=\partial_t$ is the future-directed time-like Killing vector field, $T^{(0)}$ is the background stress-energy tensor of the RN-dS solution, and 
\begin{equation} v=\frac{1}{r^2}\partial_r \qquad n=\sqrt{U}v\,,\end{equation}
is a harmonic vector field which satisfies $\nabla^av_a=0$ and which points in the outgoing radial direction. $n_a$ is the corresponding unit spacelike vector. The expansion of the  left hand side of \eq{eep} to first order in $\epsilon$ can be found in Appendix \ref{app:EEex}.

The quantum perturbation to the stress-energy tensor also has an expression in the form \eq{expandt}, 
\begin{equation}  \delta T_{ab}= \eta_1\, \xi_a \xi_b+  \eta_2\, v_av_b + \eta_3\, \xi_{(a}v_{b)} + \eta_4\, g_{ab},\label{expandt2}\end{equation}

with 
\begin{align} \eta_1&=\frac{1}{2} \left(\frac{T U+3 \mathcal{E}}{U^2}-r^4 \mathcal{S}\right),\nonumber\\\eta_2&=\frac{1}{2} r^4 \left(3 r^4 \mathcal{S} U^2-T U-\mathcal{E}\right),\nonumber\\\eta_3&=-2 r^4 \mathcal{T},\nonumber\\\eta_4&=\frac{1}{2} \left(\frac{\mathcal{E}}{U}+r^4 (-\mathcal{S}) U+T\right).\end{align}
where we have  introduced the quantities
\begin{equation}\mathcal{T}\equiv \delta  T_{ab}\xi^{a} v^{b},\quad \mathcal{E}\equiv  \delta T_{ab}\xi^a \xi^b,\quad \mathcal{S}\equiv  \delta T_{ab}v^a v^b.\end{equation}
Stress-energy conservation imposes constraints on these quantities. The linearized Einstein equations amount to $\delta_i=8\pi G \eta_i$. For $i=1,2,4$, these are a system of three coupled second-order ODE's in the radial variable that determine $\delta A,\delta B,\delta F$ to first order in the perturbation. Crucially, these equations do not involve time-derivatives; see Appendix \ref{app:EEex}. We are more interested in the equation for $\delta_3$,

\begin{equation}\frac{4 r (r \dot{M}-Q \dot{Q})}{-2 M r+Q^2-r^4+r^2}=-16\pi r^4 G \mathcal{T}.\label{eoms0}\end{equation}
Here, we have introduced the order $\epsilon$ parameters
\begin{equation}\dot{M}\equiv\left.\frac{M(\epsilon t)}{dt}\right\vert_{t=0},\quad \dot{Q}\equiv\left.\frac{Q(\epsilon t)}{dt}\right\vert_{t=0}.\end{equation}

Equation \eq{eoms0} is a first-order equation of motion for $(\dot{Q},\dot{M})$. We need a second equation to fully determine the dynamics. This is obtained by a similar perturbation of the second equation in \eq{eep}; we take the right hand side to be first-order in $\epsilon$ as well. One obtains from the two components of \eq{eep} a first-order ODE relating $\delta F(r)$ to the time-component of the current which is satisfied automatically due to energy-momentum conservation, and an equation of motion involving $\dot{Q}$,
\begin{equation}\dot{Q}=-4\pi \mathcal{J},\quad \mathcal{J}= U r^4j^av_a.\end{equation}
Here, $\mathcal{J}$ is independent of the radial variable, due to current conservation\footnote{One could imagine using a similar argument to conclude that  $\eta_3$ is constant. However, this does not work because there is an extra term coming from the first-order covariant derivative of the background stress-energy tensor, which is not present for the current since $j^{\text{classical}}=0$ in the background.}. We finally have the first order equations of motion for $M$ and $Q$,
\begin{equation}\dot{Q}=-4\pi\mathcal{J},\quad \dot{M}=-\frac{4 \pi  \left( r^5U G\mathcal{T} +Q\mathcal{J}\right)}{r}.\label{1eoms}\end{equation}
Here, $U(r)$ is evaluated on the background solution.  The method of separation of scales provides a solution to \eq{eep} which is valid to first order in $\epsilon$ and depends on the variable $t_1$ only. There might be secular perturbations for this variable, so the approximation we just carried out is only valid for $\epsilon t_1\ll 1$, or
\begin{equation} t \lesssim \frac{1}{\epsilon^2}.\label{appcontr}\end{equation}
Hamiltonian and momentum constraints are satisfied to first order in $\epsilon$ in our solution. In fact,  \eq{1eoms} is one of the momentum constraints; see Appendix \ref{app:EEex} for the other components.

One can also write \eq{1eoms} in terms of parameters measured on a local inertial reference frame. Then, $\mathcal{T}$ and $\mathcal{J}$ are replaced by quantities that depend on the microphysics, $M$, and $Q$, the local charge and momentum fluxes $\mathscr{J},\mathscr{T}$. We will do this at the location of the geodesic observer $r_g$, but again, physics is independent of this choice:
\begin{equation}\mathcal{J}=\sqrt{U(r_g)} r_g^2 j^a n_a=\sqrt{U(r_g)} r_g^2 \mathscr{J},\quad \mathcal{T}= \frac{1}{r_g^2} T_{ab} u^a n^b=\frac{\mathscr{T}}{r_g^2},\label{keke}\end{equation}
where $u^a$ is a unit future-directed vector field in the same direction as $\xi^a$. 

Finally, it is convenient to change the time variable in \eq{1eoms} to use the local time coordinate time coordinate $t_g$ of the geodesic observer at some particular value of the radial coordinate, $r_g$. This is related to the global time coordinate in \cite{Bousso:1996au} simply as
\begin{equation}\frac{dt_g}{dt}= \sqrt{U(r_g)}.\end{equation}
As a result, \eq{1eoms} become (from now on the dot now indicates time derivative with respect to $t_g$)
\begin{equation}\dot{Q}=-4\pi r_g^2\mathscr{J},\quad \dot{M}=-4\pi r^2_g \left( G\sqrt{U(r_g)} \mathscr{T}+\frac{Q}{r_g}\mathscr{J}\right).\label{2eoms}\end{equation}
Of course, physics should be independent of our choice of $r_g$.

A quantity of interest which is independent of the time coordinate is the rate of change of charge versus mass,
\begin{equation}\frac{dM}{dQ}=\frac{\dot{M}}{\dot{Q}}= G\sqrt{U(r_g)}\frac{ \mathscr{T}}{\mathscr{J}} + \frac{Q}{r_g}.\label{ww1}\end{equation}
This quantity determines the direction the black hole evolves in the $(M,Q)$ plane of Figure \ref{fig1}. The left-hand side is manifestly independent of $r_g$, while the right hand side is not. The way this works out is because stress-energy conservation relates $\mathscr{J}$ and $\mathscr{T}$. We will illustrate this for the particular case of a free scalar field $\Phi$ of mass $m$ and charge $q$; the story generalizes to free fields of any spin in a straightforward manner. For a free field, the vector $\delta T_{ab} \xi^b$ is proportional to the current $j_a$. A solution to the wave equation with frequency $\omega$ (with respect to the global time coordinate $t$) satisfies,
\begin{equation} T_{ab}\xi^b= \left( \frac{\omega}{q} +\Phi(r)\right) j_a,\end{equation}
where $\Phi(r)=-Q/r$ is the electrostatic potential \eq{A}. This can be checked explicitly via the expression for the Klein-Gordon conserved current. The contribution to $\mathscr{T}$ is
\begin{equation}\Delta \mathscr{T} = \frac{1}{G\sqrt{U(r_g)}}  \left( \frac{\omega}{q} +\Phi(r_g)\right) \Delta \mathscr{J}.\label{ww2}\end{equation}
The second term in \eq{ww2} cancels the $r_g$-dependence in \eq{ww1}, as it should. 

We now turn to understanding the detailed expressions for $\mathscr{J}$ and $\mathscr{T}$. We will start with the former. 

In de Sitter space the expansion of spacetime dilutes the particles produced by the Schwinger effect. This allows the system to have a steady current regime. This unlike Minkowski space, where current can build up \textit{ad infinitum}. The current one-form\begin{equation} j=\frac{\mathcal{J}_r}{U r^2}dr+\rho\, dt,\label{curry}\end{equation}
is conserved,
\begin{equation} d*j=0,\label{ccons}\end{equation}
and is nonvanishing in the charged black hole background since the black hole electric field pair produces charged carriers via the Schwinger effect. In a semiclassical treatment, the Schwinger effect implies the pair production of charged carriers in the vacuum,
\begin{equation} dj= 2\Gamma dr\wedge dt,\label{schw}\end{equation}
where $\Gamma$ is the ``Schwinger pair production rate'', which can be computed via instantonic methods \cite{Pioline:2005pf,Cohen:2008wz} and whose value depends on the background electric field and geometry. The factor of two accounts for the fact that particles are produced in pairs.  A mean field description such as \eq{schw} is good as long as the electric field varies slowly along space and time (by this we mean that the spatial gradients of the field are much smaller than the length scale associated to the field itself); we will relax this assumption later on on the Nariai branch.  As long as the local electric field is much larger than the background curvature, the pair production process happens essentially in flat space; the electrons in the pair are produced at a distance much smaller than the characteristic curvature, and just accelerate from there on.  As a result, we will use the flat space $\Gamma$ originally computed by Schwinger \cite{Schwinger:1951nm,Nikishov:1970br,Cohen:2008wz},
\begin{equation}\Gamma_{\text{flat-space,2d}}(m)=\frac{qE}{2\pi} e^{-\frac{\pi}{\chi}},\quad \chi\equiv\frac{m^2}{qE}.\label{schwinger2d}\end{equation}
Notice that, for $m=0$, equations \eq{schwinger2d} and \eq{schw} lead to $dj=\frac{q}{\pi} F$, which is precisely the chiral anomaly \cite{itzykson2012quantum,Harvey:2005it}. 

Equation \eq{schwinger2d} is only the two-dimensional result. In the present situation we must take into account the flux on the 2-sphere, by summing over the KK modes:
\begin{equation}\Gamma \approx \sum_{s}(2s+1)\Gamma_{\text{flat-space,2d}}\left(\sqrt{m^2+\frac{s(s+1)}{r^2}}\right) \approx 4\pi r^2 \cdot \frac{q^2E\vert E\vert}{8\pi^3} e^{-\frac{\pi}{\chi}},\label{schwinger4d}\end{equation}
where we have replaced the sum by an integral; this will be a good approximation as long as $qEr^2\gg1$. The result \eq{schwinger4d} is the well-known four-dimensional Schwinger pair production rate\footnote{The factor of two discrepancy with the classical result of Schwinger is due to the spin of a Dirac fermion as opposed to a scalar particle that we are considering here.} \cite{Cohen:2008wz}, multiplied by the area of the 2-sphere.

Following \cite{Hiscock:1990ex}, we will also impose that the current is purely ingoing along the future black hole horizon, and purely outgoing on the cosmological horizon. An outgoing component of the current at the horizon would describe particles moving ``against'' the electric field, i.e. emission of particles with opposite charge to the black hole overcoming the huge potential barrier and escaping to the cosmological horizon, or same charge particles beating the electric field and managing to cross the black hole horizon. For a superextremal particle, these rare processes will have an additional, huge tunneling barrier (we will come back to this in Section \ref{sec:conjs}), but they are actually important in some circumstances; for instance, any sub-extremal particle emitted by a flat-space black hole is bound to fall back. In situations like this, there will be a significant ingoing component, and the actual current will be smaller than our expressions below. 

Mostly everywhere on the shark fin, our expressions actually constitute an upper bound on the current, which will be close to saturation for superextremal particles. However, as we will explain below, on the Nariai limit due to the additional symmetries of the problem, there is no ingoing component even for very massive particles. We should also notice that, even when there is an ingoing component, the current never vanishes, since that would conflict with \eq{ccons}\footnote{In the flat space case, the statement is that anywhere around black hole even a sub-extremal particle has a (highly suppressed) chance of popping out of the vacuum. Thus, at any finite separation, the expectation value of the current is nonzero. The nucleation probability however decreases exponentially with the distance from the black hole, and therefore so does the current. There is no current flux at infinity, in accordance with the fact that a flat-space black hole cannot emit sub-extremal particles. In dS, however, this effect is cut off at the cosmological horizon, so there is always some current flowing.}.

 Neglecting their contribution, it follows that the current $j$ must be orthogonal to the null generator of the horizon, $n_H=-\frac{dr}{U(r)}+dt$, close to the future horizon. Current conservation \eq{ccons} now implies $\mathcal{J}_r $ is independent of position. These two properties together imply that
\begin{equation} \mathcal{J}_r= r_+^2\rho(r_+)=-r_c^2\rho(r_c).\label{kakashi}\end{equation}
Finally, \eq{curry} becomes a first-order equation for $\rho$
\begin{equation} \rho'(r)= 2\Gamma,\label{ua}\end{equation}
which we can integrate together with the boundary condition in \eq{kakashi} to obtain the expression for the current\footnote{The prefactor accounts for the definition of $Q$ \eq{adimp} in terms of the actual integer-valued current.},
\begin{equation}\mathcal{J}=\sqrt{\frac{g^2G}{4\pi \ell^2}}\frac{r_c^2\,r_+^2}{r_c^2+r_+^2} 2\int_{r_+}^{r_c}dr'\, \Gamma(r').\label{naruto0}\end{equation}
Using equation \eq{keke}, we can write down the expression for the local current measured by an inertial observed $\mathscr{J}_r$,
\begin{equation}\mathscr{J}_r=\sqrt{\frac{g^2G}{4\pi \ell^2}}\frac{2}{\sqrt{U(r)}r^2}\frac{r_c^2\,r_+^2}{r_c^2+r_+^2} \int_{r_+}^{r_c}dr'\, \Gamma(r').\label{naruto}\end{equation}
Equation \eq{naruto}, together with \eq{schwinger4d}, gives a quasistatic current to use in \eq{eom1} entirely as a function of $M$ and $Q$. It reduces to the expression in \cite{Hiscock:1990ex} for small black holes (again, modulo a factor of two due to the electron spin degeneracy).

It is important to understand the regime of validity of \eq{naruto}. This is set by backreaction on the geometry. The charge density $\rho(r)$ screens the electric field of the black hole. Its overall effect should be small. From Maxwell's equation $d*F=g^2*j$, we obtain 
\begin{equation}Q_r(r)=Q_r(r_+)+4\pi \int_{r_++\Lambda^{-1}}^rdr'\, \frac{(r')^2\rho(r')}{U(r')}.\label{sasuke}\end{equation}
The integral in \eq{sasuke} has a logarithmic divergence at the horizon, which we have regularized by introducing a regularized horizon at $r=r_++\Lambda^{-1}$, where $\Lambda^{-1}$ is some UV cutoff.  For the approximation to be trustworthy, we need 
\begin{equation}4\pi \int_{r_++\Lambda^{-1}}^rdr'\, \frac{(r')^2\rho(r')}{U(r')}\ll\sqrt{\frac{4\pi \ell^2}{g^2G}}.\label{sasuke1}\end{equation}
This will be automatically true in the quasistatic regime due to \eq{ua} and the exponential Schwinger suppression in \eq{schwinger2d}. 

Notice that in this approximation, the contribution of the Schwinger current to $\mathscr{T}$ vanishes, since the same amount of left and right-moving charge carriers are produced. Thus, we can take $\mathscr{T}$ to come entirely from Hawking radiation. At the location of the geodesic observer, this is just the associated mass flux from the two horizons taking into account the redshift factors \cite{Bousso:1996au,Bousso:1997wi,Bousso:1999iq}, 
\begin{equation}\mathscr{T}=\frac{\sigma}{(4\pi)^3}\left[r_c^2U'(r_c)-r_+^2U'(r_+)\right] ,\label{hawking}\end{equation} 
where $r_+$, $r_c$ are the black hole and cosmological horizons, and $\sigma$ is the Stefan-Botlzmann constant. Equations \eq{naruto} and \eq{hawking}, when plugged back in \eq{eom1}, completely specify the dynamics of the system as a function of the mass $m$ and charge $q$ of the charged particle.

In equation \eq{naruto} we have assumed the electric field is large compared to the background curvature, and this is a good approximation everywhere on the shark fin, except close to the neutral line, where the electric field becomes small. It turns one can work out the exact dynamics on the Nariai branch, even taking into account the curvature effects that were neglected when writing \eq{naruto}. The reason is the $dS_2\times S^2$ geometry, which allows one to translate the problem to the Schwinger effect in $dS_2$. This problem was solved exactly in \cite{Frob:2014zka}; details can be found in Appendix \ref{app:garr}. The end result is that one can write an inequality for $\mathscr{J}$,
\begin{equation}\mathscr{J}_{\text{Nariai}}=\sqrt{\frac{g^2G}{4\pi \ell^2}}\mathscr{J}_r\gtrsim\sqrt{\frac{g^2G}{4\pi \ell^2}}\frac{q^2r_c^2E\vert E\vert }{\pi^2 H_{dS_2}}\tanh\left(2\pi\frac{qE}{H_{dS_2}^2}\right)e^{-\frac{\pi}{\chi}}\label{KKmodes2}\end{equation}
which becomes tight in the large electric field limit $qE\gg H_{dS_2}^2$. We can compare with \eq{naruto} in the Nariai limit, which is also easily computed since the two integration limits coincide and the integral collapses. One obtains
\begin{equation}\mathscr{J}_{\text{Nariai, large $E$}}=2\sqrt{\frac{g^2G}{4\pi \ell^2}}\Gamma(r_c)= \sqrt{\frac{g^2G}{4\pi \ell^2}}\frac{1}{H_{dS_2}} 4\pi r_c^2 \cdot \frac{q^2E\vert E\vert}{8\pi^3} e^{-\frac{\pi}{\chi}},\end{equation}
which is exactly the same expression as \eq{KKmodes2} in the limit $qE\gg H_{dS_2}^2$. It is also important to notice that, on the extremal curve, the expression \eq{KKmodes2} is valid for any value of $M$, and it smoothly interpolates between the Hawking result for neutral particles and the Schwinger one. 

An important question we should address is whether the flow \eq{1eoms} allows for an initially sub-extremal black hole becoming superextremal.  Let us derive a simple criterion on the flow such that this is satisfied. In other words, we want to understand under which conditions the flow \eq{1eoms} maps the region inside the blue curve in Figure \ref{fig1} to itself. Since the allowed region is defined by $\Delta>0$ in \eq{discr}, we require that the $(\dot{M},\dot{Q})$ vector along the boundary $\Delta=0$ points towards the gradient of $\Delta(M,Q)$, i.e.
\begin{equation}\dot{Q} (\partial_Q\Delta)_{\Delta=0}+\dot{M} (\partial_M\Delta)_{\Delta=0}\geq0.\label{uwuwwu}\end{equation}
Now, all over the subextremal region $(M,Q)$ plane except on the $Q=0$ axis, we expect $\dot{Q}<0$ in a theory with charged particles, since the black hole horizon always tends to discharge and this only stops at $Q=0$. 

We separately discuss constraint (\ref{uwuwwu}) near the extremal branch and near the Nariai branch. A useful relation, valid all over the edge of the sharkfin ($\Delta=0$) is
\begin{equation}
\left(\frac{\partial_Q\Delta}{\partial_M\Delta}\right)_{\Delta=0} = -\frac{Q}{r}
\end{equation}
with $r$ the horizon radius. On the Nariai branch one  exactly saturates inequality (\ref{uwuwwu}). For that we use the vanishing of the $\sqrt{U}_g$ term in (\ref{ww1}) together with the relation between $r_g$ and $Q$ as given in (\ref{rc2d}).   For the extremal branch, Figure \ref{fig1} shows that $(\partial_Q\Delta)_{\Delta=0}<0$ whereas $(\partial_M\Delta)_{\Delta=0}>0$. Equation (\ref{uwuwwu}) is satisfied because $\mathscr{T}<0$ at the horizon. This is because the black hole is colder than the cosmological horizon, resulting in a net mass gain which dominates over Schwinger radiation in the quasi-static regime. 

\begin{figure}
\begin{center}
\includegraphics[width=0.47\textwidth]{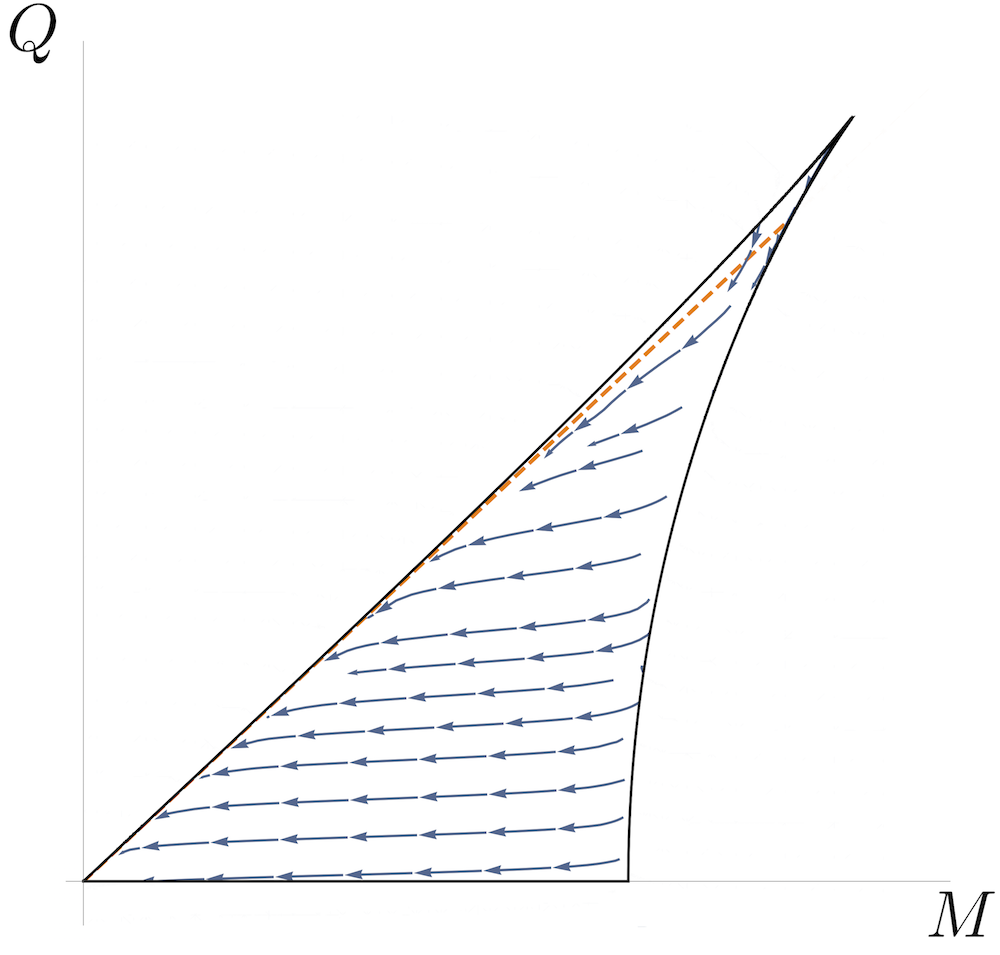}
\vspace{0.5cm}

\caption{\small{Schematic representation of the quasistatic flow on the shark fin generated by the combination of the Schwinger effect and Hawking radiation. Black hole solutions evolve towards the lukewarm or charged Nariai lines, losing charge very slowly in the meantime. Eventually, they evaporate completely. The flow stays completely within the shark fin, the black holes never turning superextremal. The flow follows exactly the charged Nariai line, so ignoring quantum fluctuations charged Nariai black holes remain Nariai while slowly discharging to the neutral solution. For illustration purposes, we took $\mathscr{J}\sim 1$ and $\mathscr{T}\sim  M-Q$ to draw the plot.}}\label{shark-flow}
\end{center}
\end{figure}

Figure \ref{shark-flow} shows a schematic plot of the flow generated by the solutions in the shark fin, where we have taken $\mathcal{J}$ and $\mathcal{T}$ to be constant\footnote{Similar plots can be found for the Kerr case in \cite{Anninos:2010gh}.}. Our results are of course not exact; the detailed expressions are obtained by performing the integral in \eq{naruto}, which can be evaluated explicitly in terms of error functions \cite{Hiscock:1990ex}. Still, Figure \ref{shark-flow} gives a roughly correct schematic picture of the flow of black hole solutions on the shark fin: Solutions flow to either the lukewarm line or the Nariai branch, and then the neutral line. All solutions evaporate completely, eventually. Notice that this conclusion is insensitive to the precise values of $(m,q)$, as long as the Schwinger effect is exponentially suppressed. This has consequences for the WGC in dS, which we discuss in Section \ref{sec:conjs}.

Finally, a couple comments. It is worth emphasizing that we have assumed steady-state current. This is a physical assumption, which can be relaxed. One could for instance start with a black hole with no current present at all initially. But then the current will grow and, over the period of a Hubble time, such a state will develop into the steady current state.

We are looking at a quantum effects in a black hole background; we should make sure that there are no classical instabilities present. For instance, the Schwinger effect for near-extremal black holes in $AdS$ actually takes the form of a classical instability \cite{Crisford:2017gsb}, and a charged scalar field on the RN-dS is known to have classical instabilities for some values of $M,Q,m,q$ \cite{Dias:2018ufh}. The instability discussed in these references only takes place for very light fields, away from the quasistatic regime, and for small values of the charge. As illustrated in Figure 1 of \cite{Dias:2018ufh}, the instability seems not to be relevant near the charged Nariai branch. 

We should also point out\footnote{We thank T. Banks for bringing this up.} that the quasistatic computation we just described fixes the black hole at the center of the static patch -- in other words, it ``follows'' the black hole along its worldline. This worldline is not a geodesic. The black hole receives radiation from the cosmological horizon and emits some itself, exchanging energy and momentum, so there will be some sort of Brownian motion \cite{BanksMadrid,BanksMadrid2}.  We can estimate the time that it takes for the geodesic observer to be kicked out when the black hole is much smaller than the Nariai limit\footnote{To carry out a similar analysis for a very large black hole is beyond the scope of this paper.}. Consider a small black hole on the lukewarm line. Its center of mass degrees of freedom will have thermalized and so it will have a mean kinetic energy of $3/2 T$. In empty $dS$ space, a free particle of mass $M$ and energy $E$ takes a time 
\begin{equation}t_{\epsilon}=\sqrt{\frac{E^2}{E^2-M^2}}\arctan(1-\epsilon)\end{equation}
to reach $r=1-\epsilon$ the cosmological horizon. Taking $T=H/2\pi$ to be the de Sitter temperature, one gets a timescale suppressed as $t_{\epsilon}\sim \sqrt{M/H}$. This is a long time since black holes are heavier than a Planck mass, but not exponentially suppressed, and so deep into the quasistatic regime a true geodesic observer will thus see the black hole kicked out of her static patch by the radiation, and will never see the evaporation come to an end. One could try to fix this by always sticking to the black hole center of mass, although this will no longer be described by the usual static patch Hamiltonian. In any case, we think these subtleties are not likely to affect our subsequent discussion, since most of our arguments in Section \ref{sec:conjs} will be about large black holes (close to the Nariai limit), for which the simple picture above does not hold, and also in the adiabatic regime, where the black hole discharges quickly.

We must also discuss Bousso's beading process \cite{Bousso:1998bn,Bousso:1999ms,Bousso:2002fq}. These references describe a family of classical instabilities for the full four-dimensional charged Nariai geometry, coming from inhomogeneous fluctuations of size of the $S^2$ on the Nariai geometry. Where the radius of the $S^2$ becomes smaller than the Nariai value, the geometry collapses to a black hole; where it is slightly larger, it behaves like the exterior of a $dS_4$ geometry, with the spheres becoming asymptotically large in the asymptotic future. In short, these are fluctuations that drive the black hole towards sub-extremality. On the exact Nariai geometry, a perturbation in the $n$-th spherical harmonic leads to the formation of ``beads'', spacetimes containing $n$ near-extremal RN-dS black holes.

Interestingly, the dynamics of this process is very different below or above the intersection of the lukewarm line with the charged Nariai line; above, the positive heat capacity of the black hole makes the Nariai geometry stable locally, so quantum fluctuations lead to an endless beading process and a fractal structure in the asymptotic future. Below the lukewarm line, at most a finite number of beads can be produced (how many exactly depends on the particular mode that dominates the perturbation), and the geometry in any given static patch goes back to a slightly sub-extremal black hole picture. 
 
These predictions were computed with a one-loop quantum effective action which so far has always agreed with the thermodynamic results whenever available \cite{Bousso:1998bn,Bousso:1999ms,Bousso:2002fq, Niemeyer:2000nq}. For instance, while it seemed that neutral Nariai black holes could sometimes initially anti-evaporate \cite{Bousso:1997wi}, in the long run their behavior was always dictated by the semiclassical model. Effectively, what this means is that, below the lukewarm line, one should expect quantum effects to push the solution slightly off the Nariai branch, towards sub-extremality, where Hawking radiation drives evolution towards the lukewarm branch which then discharges quasi-statically, as in Figure \ref{shark-flow}. Above the lukewarm line one really stays in the Nariai branch for a long time, slowly discharging (and beading along the way) until crossing the lukewarm threshold.

\subsection{Adiabatic discharge:  \texorpdfstring{$m^2\ll qE$}{m2<qE}}
We are also interested in quantifying how quickly the relaxation of the system happens when the mass of the charged carriers is below $qE$. Working again in the limit $qE\gg H^2$, the discharge effectively takes place in flat space, and on very short times compared to the Hubble scale. The electric field is so strong that it is immediately screened  out locally by the pair-produced charge carriers, which subsequently annihilate each other, leading to radiation. As a result, one can model this kind of evaporation by replacing the electric field and charge matter at $t=0$ by pure radiation with the same energy density. 

We will only work out the detailed dynamics on the charged Nariai branch. Since the geometry is $dS_2\times S^2$, one can work in the effective two-dimensional theory, which greatly simplifies the problem. We will start with the 4d Einstein-Maxwell action
\begin{equation}S=\int d^4x\sqrt{-g}\left[ \frac{1}{16\pi G}\left[-R+\frac{6}{\ell^2}\right]+\mathcal{L}\right].\label{rasengan}\end{equation}
and set $\ell=1$. $\mathcal{L}$ describes the matter (electromagnetic fields + charged matter) we must add to the system. We will now work out the dynamics . To do this, we will work on a $S^2\times \mathbb{R}^2$ ansatz. We will also follow \cite{Frob:2014zka} and resort to a FRWL\footnote{The L stands for Lema\^{i}tre. It must be included in any Leuven paper that discusses this metric.} open slicing which is smooth beyond the horizon, with coordinates $x,\tau$ such that the four-dimensional metric is
\begin{equation} ds^2=e^{-\phi}(-d\tau^2+ a^2(\tau)dx^2)+ e^{2\phi}d\Omega^2.\label{2dmetr}\end{equation}
Plugging this back into \eq{rasengan}, the kinetic term for $\phi$ becomes a total derivative and drops out, as usual in two-dimensional gravity models. The resulting effective action is
\begin{equation} S=\frac{1}{8\pi G} \int d\Omega_2 dV_2\left\{ e^{\phi} (\mathcal{L} +3)-\frac{1}{2}\left[2e^{-\phi}+e^{2\phi} R_{g_2}\right]\right\}.\label{metricansatz}\end{equation}
where $g_2$ is the two-dimensional metric $d\tilde{s}^2=-d\tau^2+a(\tau)^2 dx^2$. Using $dV_2= a(\tau)d\tau\wedge dx$, and
\begin{equation}R_{g_2}=2 \frac{\ddot{a}}{a},\end{equation}
one gets, after integration by parts,
\begin{equation}8\pi G\, S=\int d\Omega_2 d\tau dx\left\{\dot{(e^{2\phi})} \dot{a} -e^{-\phi}a+3e^{\phi} a\right\}+8\pi G \int dV_2d\Omega_2 \sqrt{-g}\mathcal{L}.\label{act0}\end{equation}
The 4d stress-energy tensor is 
\begin{equation} T_{\mu\nu}\equiv \frac{2}{\sqrt{-g}}\frac{\delta S_M}{\delta g^{\mu\nu}},\end{equation}
and so, the variation of the matter part of the action (the last term in \eq{act0}) is
\begin{equation} 8\pi G \delta S_{\text{mat.}}=4\pi G\int d\tau dx\,   a e^{\phi} T_{\mu\nu} \delta g^{\mu\nu},\end{equation}
in terms of the four-dimensional stress-energy tensor $T_{\mu\nu}$. Since 
\begin{equation}\delta g^{\mu\nu}=-2\frac{e^{\phi}}{a^3}\delta^{\mu,x}\delta^{\nu,x}\delta a+ \delta\phi\left[3e^{\phi}g_2^{\mu\nu}-2g^{\mu\nu}\right],\end{equation}
the equations of motion are 
\begin{align}\ddot{(r^2)}&=3r-\frac{1}{r}-8\pi G \frac{r^2}{a^2} T_{xx},\label{eom1}\\ 2r^2\frac{\ddot{a}}{a}&=\frac{1}{r}+3r+4\pi G r\left[ 3T_2-2T\right],\label{eom2}\end{align}
where we have introduced the two-dimensional trace $T_2\equiv e^{\phi} T_{\mu\nu} g^{\mu\nu}_2$. Finally, in the problem under consideration, the evolution is governed by electromagnetic stress-energy, which is traceless. We will decompose the traceless stress energy tensor of radiation as
\begin{equation}T_{\mu\nu}= \rho u_{\mu} u_{\nu} + p_1 \gamma_{\mu}\gamma_{\nu} + p_2 g_{\mu\nu}^{\Omega},\end{equation}
where $u^{\mu}$ is a time-like future-pointing unit norm vector , $\gamma_\mu$ is the same thing on the $x$ direction, and $g_{\mu\nu}$ is just the angular part of the metric \eq{2dmetr}. Tracelessness implies $\rho=p_1+2p_2$.  In terms of these quantities, \eq{eom1} and \eq{eom2} become
\begin{align}\ddot{(r^2)}&=3r-\frac{1}{r}-8\pi G r p_1,\label{eom3}\\ 2r^2\frac{\ddot{a}}{a}&=\frac{1}{r}+3r+3r\left[8\pi G\frac{p_1-\rho}{2}\right].\label{eom4}\end{align}

Different electromagnetic fields correspond to different $\rho,p_1,p_2$. For instance, in the quasistatic regime discussed in the previous subsection, the electric field is constant,  which means $\rho=-p_1=p_2$. Also, in this regime we can neglect the backreaction of the matter, the time derivatives of $\phi$ are small, and those of $a$ are basically constant, so that $r \ddot{a}/a=H^2$. Then, taking
\begin{equation}\mathcal{L}= \frac{1}{4g^2} F_{\mu\nu}F^{\mu\nu}\end{equation}
as in Section \ref{sec:pre}, together with the electric field
\begin{equation} F= \frac{g^2}{4\pi}\frac{Q_r}{r^3} a\, dx\wedge d\tau,\end{equation}
one obtains (after using \eq{adimp}) that
\begin{equation}-8\pi G \rho= 8\pi G p_1= 8\pi G \frac{p_1-\rho}{2}= -\frac{Q^2}{r^4}.\label{instrr}\end{equation}
Plugging back in \eq{eom3} and \eq{eom4}, one obtains
\begin{align}\ddot{(r^2)}&=3r-\frac{1}{r}+\frac{Q^2}{r^3},\label{eom5}\\ 2r\frac{\ddot{a}}{a}&=\frac{1}{r^2}+3\left(1-\frac{Q^2}{r^4}\right).\label{eom6}\end{align}
Upon replacing $r\ddot{a}/a$ by $H^2$, and $\ddot{r}$ by 0, \eq{eom5} and \eq{eom6} become two algebraic equations,
\begin{equation} 3r^4-r^2+Q^2=0,\quad 2H^2=\frac{1}{r^2}+3\left(1-\frac{Q^2}{r^4}\right).\label{adiabb}\end{equation}
These are the same evolution equations which one obtains from the quasistatic evolution for the full 4d geometry, particularized to the Nariai branch. In particular, the solution to the first equation is \eq{rc2d} and then substituting in the second one gets \eq{hubble2d}. Then, the slow time evolution dictated by the slow Schwinger discharge as described by the first equation in \eq{2eoms}.

 As mentioned above, we will model the adiabatic regime by replacing the electric field and charge matter at $t=0$ by pure radiation with the same energy. That is, one still takes $\rho$ as in \eq{instrr}, but $p_1$ and $p_2$ are no longer related as for a constant electric field. We will parametrize
\begin{equation}p_1=\alpha\rho, p_2=\beta\rho,\quad \alpha+2\beta=1.\end{equation}
 $\alpha$ and $\beta$ are both positive, since we are now working with radiation, and not with electrostatic fields as before. The precise relation between these two is dictated by the physics of the annihilation process and immediately after. The nucleated pairs are produced and then accelerated for a mean free path, after which they annihilate. Since the charge carrier is massless, and the electric field is very string, it mediates very efficient interactions and it is perhaps reasonable to assume that the system thermalizes locally. In that case, it seems reasonable to take $\beta=\alpha=1/3$, corresponding to the stress-energy tensor of a four-dimensional CFT in the thermal state \footnote{In the opposite limit, where no thermalization occurs, one could compute $\alpha,\beta$ by integrating the angular distribution of photons produced during the scattering fo charged carriers. The perturbative QED cross section for $e^+e^-\rightarrow\gamma\gamma$ is  heavily peaked as usual in the forward direction, when the t-channel goes on-shell. Hence it would seem reasonable to take the radiation to be forward-pointing, $\beta\approx0$, but we will keep the analysis general. Both $\alpha,\beta$ are constants during time evolution if one neglects scattering of photons with the background curvature or with each other.}. We will focus on this case from now on, although the precise values of $\alpha,\beta$ won't affect our conclusions.  Conservation of stress-energy then fixes
\begin{equation}\dot{\rho}+\rho\left(2\dot{\phi}+\frac{4}{3}\frac{\dot{a}}{a}\right)=0,\end{equation}
which integrates to
\begin{equation}\rho=\rho_0e^{-2\phi}a^{-4/3}.\label{e0}\end{equation} 
We also have (in Planck units already)
\begin{equation} \rho_0=\frac{Q_0^2}{r_0^2}\label{eof4}\end{equation}
due to energy conservation, where $Q,r_0$ are the initial charge and radius (related to $Q_0$ as in \eq{adimp}). 
Introducing the convenient variable $s\equiv r^2$, the equations of motion \eq{eom3}, \eq{eom4} become, after using \eq{e0},
\begin{align}\ddot{s}&=3\sqrt{s}-\frac{1}{\sqrt{s}}\left(1+\frac{\rho_0}{a^{4/3}}\right),\label{eom7}\\ \frac{\ddot{a}}{a}&=\frac{1}{2s^{3/2}}\left(1-\frac{\rho_0}{a^{4/3}}+3s\right),\label{eom8}\end{align}
with the initial condition $a_0=1$. 

\begin{figure}[!htb]
\begin{center}
\includegraphics[width=0.47\textwidth]{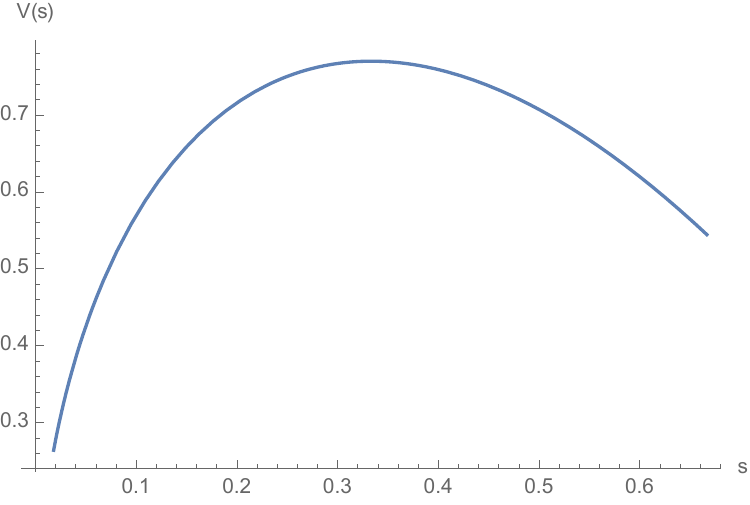}
\vspace{0.6cm}
\caption{\small{Plot of the effective potential $V(s)\equiv2\sqrt{s}(1-s)$. The maximum at $s=1/3$ corresponds to the neutral Nariai value. Since the value of $s$ of the neutral Nariai black hole is higher than that of any of its charged counterparts,  we are left with an initial condition to the left of $s=1/3$ after the electric field discharges instantaneously. The equation of motion \eq{eom7} then implies one is pushed even further to the left, towards $s=0$.}}\label{fig6}
\end{center}
\end{figure}

Since $\rho_0<1$ (this is equivalent, due to \eq{eof4}, to $Q_0^2/r_0^2<1$, which is true everywhere on the Nariai branch) the second equation of motion tells us that the expansion is always accelerated. In fact, it is Friedmann's equation in two dimensions. The first tells us that $\dot{s}<0$, so that the size of the two-sphere decreases. After a while, one can ignore the $\rho_0$ term in \eq{eom7} and \eq{eom8}, since the redshift has diluted the radiation away. Equation \eq{eom7} can then be integrated to yield an effective potential for $s$,

\begin{equation} \frac{1}{2}\dot{s}^2+2\sqrt{s}(1-s) = E\end{equation}
The potential is depicted in Figure \ref{fig6}. The maximum happens at $s=1/3$, which is the size of the 2-sphere for a neutral Nariai black hole. The initial condition is therefore to the left of this point, and moving to the left. As one approaches $s\sim0$, the scale factor blows up; the horizon size becomes tiny, and the curvatures become Planckian. Thus, this is a Big Crunch. In fact, it is a crunch of a very similar nature to the one encountered as one approaches the spacelike singularity inside a Schwarzschild-dS black hole; the 2-spheres go to zero size, and the resulting 2d cosmological constant obtained after dimensional reduction from the 4d cosmological constant becomes huge as well. 

\begin{figure}[!htb]
\begin{center}
\includegraphics[width=0.47\textwidth]{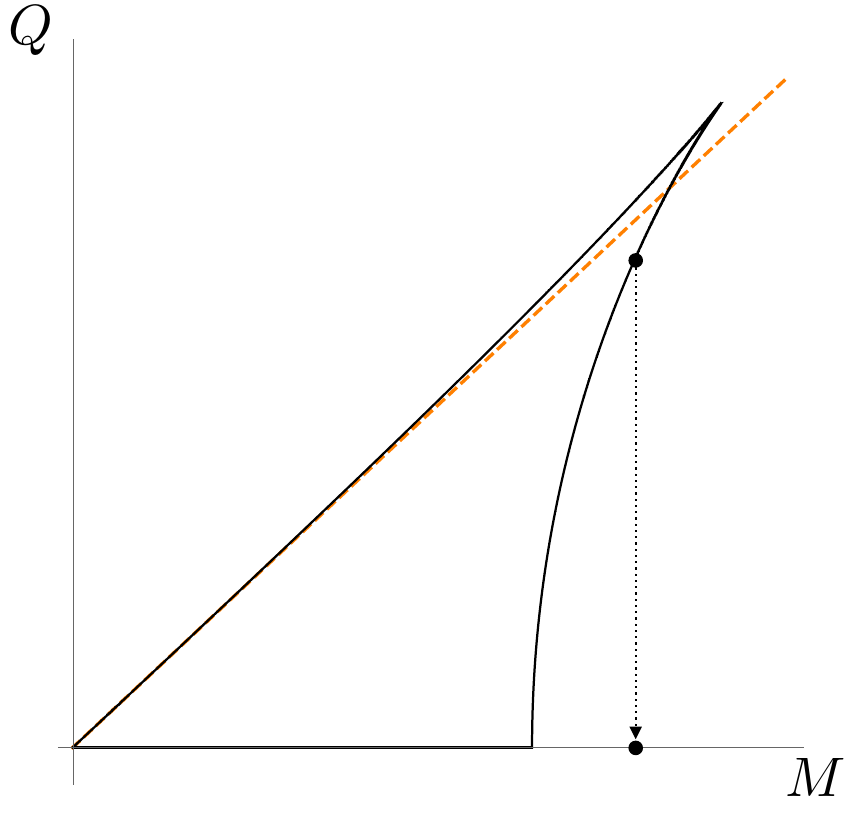}
\vspace{0.6cm}
\caption{\small{Nariai black hole evaporation in the adiabatic limit of a massless charge carrier. The electric field is screened very quickly, getting replaced by radiation, and so the black hole loses its charge without losing mass. As a result, we end in a ``superextremal'' neutral Nariai geometry (plus radiation), which ends in a big crunch.}}\label{fig6ymedio}
\end{center}
\end{figure}

What is happening is that, by having all the electric field discharge instantaneously, we find ourselves in a situation with no charge, but a mass above that of a neutral Nariai black hole. Effectively, we have made a transition to a ``superextremal'' Nariai solution, as illustrated in Figure \ref{fig6ymedio}. These solutions are cosmological, describing either a Big Bang or a Big Crunch where the 2-spheres go to zero size. These correspond to the left or right branches of the potential in Figure \ref{fig6}. As we go from slightly sub-extremal to slightly super-extremal mass, the conformal diagram of the spacetime first looks like $dS_2\times S^2$, and at super-extremality it disconnects into two pieces. The one describing what used to be the original static patch has been devoured by the black hole and in a sense describes the ``black hole interior'', with a crunch singularity in the future. The other one describes what used to be the region of global $dS$ outside of the static patch of the black hole, and the two-spheres grow asymptotically to empty $dS$.

Starting with an extremal charged Nariai black hole whose electric field disappears adiabatically, we find a configuration that asymptotes to the same crunch that the super-extremal Nariai geometry. After the transition, we find ourselves in the crunching portion of the superextremal Nariai spacetime, plus some radiation that quickly redshifts. Although we have not proven this, one would expect the time for the beading process or other quantum effects to be irrelevant in this context, as the dominant source of stress-energy is the radiation, and the crunch happens quickly, within a Hubble time (see Appendix \ref{app:qadiab}). 

Since this crunch will be crucial for our argument below, we must make sure it is not an artifact of the Nariai approximation\footnote{We thank Grant Remmen and J. L. F. Barb\'{o}n for discussions on these points.}. Is the crunch somehow avoided if there are small inhomogeneities (that could collapse into black holes and then leave the static patch), or if we start slightly off the Nariai branch, so that the size of the two-spheres is not exactly constant? What about the choice of initial hypersurface? The choice of quantum state is also important, see e.g. \cite{Aalsma:2019rpt}. Following \cite{Frob:2014zka}, we chose an open FRW slicing, in which the dynamics is homogeneous in the Nariai limit. This is convenient for our calculations but it is important to check that the crunch is actually independent of this choice.  We could have picked any other spacelike surface, like the $t=0$ hypersurface in the static patch. This would map to inhomogeneous radiation initial data for some later $\tau=\text{const.}$ hypersurface in the FRW ansatz \eq{2dmetr}. 

These concerns are averted by the fact that the crunch is somehow ``local'' -- it happens for each observer and it is not very much affected by energy inflows or outflows, as long as these are small enough--. The precise way to state this is that the crunch is guaranteed by GR singularity theorems\footnote{We thank Raphael Bousso for pointing us in the right direction here.}, as we explain in Appendix \ref{app:crunch}.

\section{Black hole decay in dS and an EFT constraint}\label{sec:conjs}
We have now finished presenting the technical results in this paper. The formalism developed in these previous Sections provides a means to understand how black holes evaporate in de Sitter space. This is a calculation within in General Relativity which, to our knowledge, has not appeared in the literature before.

By contrast, in the present Section we will indulge in hopefully interesting speculation. We wish to emulate some of the logic that leads to WGC in flat space or AdS, to hopefully learn something about the properties of effective field theories in (quasi-)de Sitter spaces. 

\subsection{Strong gravity in de Sitter space}\label{sec:sgc}

In Section \ref{sec:decay},  we have seen that all black holes evaporate and decay back to empty de Sitter space, which is consistent with the interpretation of de Sitter as a thermal equilibration process. There is just one exception: Black holes close to the Nariai branch, whenever there is at least one particle in the adiabatic regime $m^2\ll qE$. In this case, the geometry evolves towards a crunch, and never returns to empty de Sitter space. The adiabatic evolution forces a black hole outside of the shark fin; since the Nariai branch tilts to the right in the $(M,Q)$ plane, losing the charge without losing the mass leads to a neutral solution with a mass higher than that of the neutral Nariai black hole. 

This is also problematic because once we land in a point outside the shark fin, we are free to deform the solution (say on a fixed Cauchy slice) away from this, towards positive charge or arbitrarily large mass. The latter is hard to reconcile with the picture of de Sitter space as having a finite-dimensional Hilbert space \cite{Banks:2000fe,Banks:2003cg,Banks:2005bm,Witten:2001kn}. Furthermore, charged superextremal solutions are continuously connected to neutral ones, since one can go around the tip of the shark fin in Figure \ref{fig1}. 

These are the superextremal black holes with naked singularities  that are usually forbidden by Cosmic Censorship\footnote{Cosmic Censorship is no fundamental principle. Solutions with unbounded curvature but with a consistent UV description are familiar in String Theory, e.g. orbifolds, KK monopoles, or all sorts of branes \cite{ibannez2012string}. Gubser's criterion \cite{Gubser:2000nd} explains that admissible singular solutions with vanishing temperature must always be cloaked behind a horizon when an arbitrarily small temperature is turned on. Superextremal RN black holes would not satisfy this.}, in its loose original meaning \cite{Wald:1997wa} that Einstein's theory should not develop arbitrarily large curvatures which are not shielded by a horizon.  In flat space, existence of a weak gravity particle ensures that large extremal black holes can decay without becoming superextremal. As usual, this connection can be made far more precise in AdS, where a series of papers \cite{Crisford:2017zpi,Crisford:2017gsb,Horowitz:2019eum} have established that certain violations of Cosmic Censorship are avoided if a version of the WGC is satisfied. 

There is also a lot of evidence that physics in the static patch of dS looks very much like that of an ordinary thermal system \cite{Banks:2000fe,Spradlin:2001pw,Banks:2003cg}: There is a thermal bath coming from the horizon at a temperature $(2\pi\ell)^{-1}$, and whatever initial perturbation one considers, it tends to be ``smoothed out'' by inflation, i.e. features fall behind the cosmological horizon and disappear. In these processes the generalized second law of thermodynamics is always satisfied \cite{Anninos:2012qw}. So it is not unreasonable to regard dynamics in de Sitter in general as describing thermal relaxation, where perturbations equilibrate and disappear after a certain time. This picture is exactly what one gets in the quasistatic regime, where black holes decay slowly back to empty dS; it fails spectacularly in the adiabatic regime. 

So, a theory with charged extremal black holes and particles in the adiabatic regime is hard to reconcile with a self-contained description of the physics in the static patch in terms of a finite-dimensional Hilbert space. If one wants to make both compatible, we only see three possibilities:
\begin{itemize}  
\item The Big Crunch at the end of the black hole collapse transitions back to empty dS space via some magical quantum gravity process.
\item Charged black hole solutions are not allowed in a theory with particles in the adiabatic regime.
\item Particles in the adiabatic regime are not allowed, and this constitutes a Swampland-type constraint.
\end{itemize}
We can say little about the first possibility. The second seems difficult at least in the semiclassical regime, because again, we can smoothly deform empty de Sitter space into any other solution in the charged Nariai branch\footnote{In absence of charged particles, one can construct gravitational instantons that mediate transitions between empty de Sitter space and the Nariai or lukewarm, and which furthermore enforce detailed balance \cite{Ginsparg:1982rs,Bousso:2002fq,Dias:2004px}. However, it is unclear whether these tunneling transitions are destabilized by the presence of light charged modes. In any case, the argument that the Nariai solutions are smoothly connected to the vacuum in configuration space still stands.}. So we will entertain the third possibility. 

If indeed the weird properties of the adiabatic regime -- lack of thermalization, failure of detailed balance, seemingly huge Hilbert space -- are a problem, perhaps we should avoid it altogether. That means $m^2>qE$ for every particle and on every point on the charged Nariai branch. The electric field is largest at the ultracold point, where $E=\sqrt{6} M_P g H$ and we have used $M_P^2=(8\pi G)^{-1}$. Thus, avoiding the adiabatic regime is equivalent to imposing\footnote{We write $\gtrsim$ because we have only analyzed the two extreme regimes where $m^2$ is very small or very large compared to $qE$. Due to the exponential prefactor in \eq{schw00} the quasi-static regime is reached after $m^2/qE$ is of order 10. The precise threshold can be computed by demanding that the excess charge \eq{sasuke} is small (say $10^{-5}$) compared to the background. We write this footnote thinking of all our friends that dislike symbols like ``$\gtrsim$''.}
\begin{equation}m^2\gtrsim q\, g M_P H,\quad\text{for every particle in the spectrum}.\label{sgc}\end{equation}
We consider \eq{sgc} a potential candidate for a Swampland constraint in de Sitter. As we tried to emphasize, it is derived via similar heuristics to the ones that lead WGC in flat space. Unlike WGC, however, we lack any stringy examples to check against \eq{sgc}, so this result is definitely on shakier ground. Equation \eq{sgc} is just a consistency condition so that the thermal picture of de Sitter can hold. 

In fact, there are good reasons to at least entertain the possibility that there are no metastable de Sitter vacua in string theory \cite{Danielsson:2018ztv,Obied:2018sgi,Agrawal:2018own}. If supersymmetric string theory as we know it today really describes the real world, the logical conclusion would be we are in some sort of quintessence scenario, so it makes sense to ask what would happen to \eq{sgc} in such a case. This analysis is carried out in Appendix \ref{app:qadiab}; one finds that quintessence does not change the conclusion, and that violating  \eq{sgc} still leads to a crunching spacetime.

We now make a couple of comments before crudely discussing the phenomenological implications.  First, the constraint becomes trivial as $H\rightarrow0$, as it should since it is a pure de Sitter phenomenon. On the other hand it is nontrivial and in particular forbids all charged particles if we take $M_P\rightarrow\infty$ at constant $H$ and $g$. This suggests that it might not be possible to consistently decouple gravity in dS\footnote{We do not claim that QFT makes no sense in a dS background; it does, at least perturbatively. We just mean that, given a gravitational model with finite $M_P$, there might be no way to ``deform'' it, either continuously or by small discrete steps, to one where gravity is absent.}. In a more holographic language, this means that Einstein\footnote{In particular, higher-spin theories are not Einstein gravity and the discussion in this paper does not apply to them. Higher-spin theories in de Sitter are conjectured to have holographic duals \cite{Anninos:2011ui} which do possess a large $N$ limit.} dS holographic models, if they exist, should not fall in large $N$ families where the low-energy sector ``stabilizes'' and we can make gravity as weak as we want by taking large $N$. Rather, they should be ``sporadic'' or isolated theories. This point of view was also advocated in \cite{Witten:2001kn}.  Second, we emphasize the statement holds for every particle in the theory, since just one violating the bound would be enough to cause an adiabatic discharge.  This is unlike Weak Gravity, where just a single particle is enough.  This difference  reflects that Weak Gravity is about ensuring that black holes decay, while \eq{sgc} is about ensuring they don't do it in a particular way.  Extremal black holes satisfy \eq{sgc} as well. Third, one should only apply this constraint in situations where a Nariai black hole makes sense. In particular it does not apply to Higgsed or confined gauge fields. 

There is also an interesting interplay between \eq{sgc} and magnetic/sublattice versions of the WGC which predict an upper bound on the cutoff scale of quantum gravity \cite{Heidenreich:2017sim}. Given a mass for a charge carrier, \eq{sgc} predicts an upper bound on the gauge coupling in terms of the mass of the lightest charged carrier,
\begin{equation}g\lesssim \frac{m^2}{q M_P H},\end{equation}
which translates to an upper bound on the cutoff of the EFT. 

Throughout this work, we focus on the case with a single species of charged particle. Of course, if one has multiple species of charged particles, they will all contribute to the black hole decay. This allows one to trigger the adiabatic regime, even if the individual particle species of particles are massive. However, the annihilation of the produced particles into neutral radiation may be slower than in the massless case. For instance, if one has order $\exp(m^2 / q E)$ particle species of mass $m$ and charge $q$, this compensates for the exponential suppression of Schwinger pair production. Demanding the adiabatic regime does not occur then bounds the number of charged particle species at certain mass in such cases. This has in fact interesting implications, which we discuss in Subsection \ref{sec:wgc}.
\subsubsection{Phenomenological implications}
The nice thing about making Swamp-like statements in Sitter space is that one may compare them to the real world. Taking the $U(1)$ to be electromagnetism, the scale on the right hand side of \eq{sgc} is 
\begin{equation} \sqrt{g M_P H}\sim 10^{-3}\, eV,\label{cjsc}\end{equation}
or around the vacuum energy density scale/neutrino mass scale. The Standard Model satisfies our constraint \eq{sgc}, since the lightest electrically charged particle is the electron, eight orders of magnitude above \eq{cjsc}.

\begin{figure}
\begin{center}
\resizebox{0.7\textwidth}{!}{\begin{tikzpicture}
  \draw[line width=0.12mm,black] (-1,0)  -- (1,0);
\node [centered,scale=0.2] at (-1,-0.09) {$H$};
\draw[fill] (-1,0) circle [radius=.5pt];
\node [centered,scale=0.2] at (-0.016414,-0.09) {$\sqrt{ g M_PH}$};
\draw[fill] (-0.016414,0) circle [radius=.5pt];
\node [centered,scale=0.2] at (1,-0.09) {$M_P$};
\draw[fill] (1,0) circle [radius=.5pt];
 \draw[fill,blue] (0.218859,0.2) circle [radius=.2pt];
 \draw[line width=0.05mm,blue] (0.218859,0.2)  -- (0.218859,0);
 \node [above left,scale=0.2] at (0.218859,0.2) {$e$};
 
  \draw[fill,blue] (0.322428,0.3) circle [radius=.2pt];
 \draw[line width=0.05mm,blue] (0.322428,0.3)  -- (0.322428,0);
 \node [above left,scale=0.2] at (0.332428,0.28) {$p$};
 
  \draw[fill,blue] (0.383575,0.4) circle [radius=.2pt];
 \draw[line width=0.05mm,blue] (0.383575,0.4)  -- (0.383575,0);
 \node [above,scale=0.2] at (0.383575,0.4) {$W$};
 \draw[fill,blue] (0.292337,-0.2) circle [radius=.2pt];
 \draw[line width=0.05mm,blue] (0.292337,-0.2)  -- (0.292337,0);
 \node [below left,scale=0.2] at (0.292337,-0.2) {$\mu$};
 
  \draw[fill,blue] (0.348545,-0.3) circle [radius=.2pt];
 \draw[line width=0.05mm,blue] (0.348545,-0.3)  -- (0.348545,0);
 \node [below left,scale=0.2] at (0.348545,-0.3) {$B$};
 
  \draw[fill,blue] (0.394122,-0.4) circle [radius=.2pt];
 \draw[line width=0.05mm,blue] (0.394122,-0.4)  -- (0.394122,0);
 \node [below left,scale=0.2] at (0.394122,-0.4) {$t$};
  \end{tikzpicture}}
\vspace{0.5cm}

\caption{\small{Logarithmic scale of energies displaying the Hubble and Planck scales on both ends, the masses of some SM electrically charged particles, the Higgs vev scale, and the new scale $\sqrt{g M_P H}$ in \eq{cjsc}. The constraint \eq{sgc} implies that all charged particles should lie to the right of this scale, which is satisfied.}}\label{fig4}
\end{center}
\end{figure}
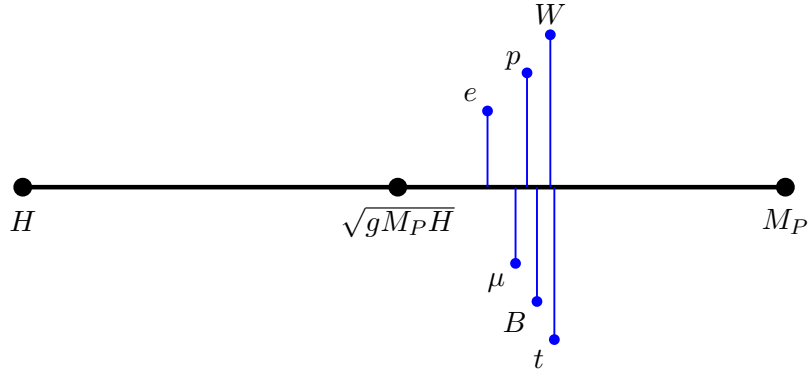

While at first sight it seems that \eq{sgc} is satisfied with room to spare, one should remember the immense hierarchy between Planck and Hubble scales, of sixty orders of magnitude.  \eq{sgc} predicts a new scale which is the geometric mean of these two, and the eight orders of magnitude in the electron should be compared with these sixty, as illustrated in Figure \ref{fig4}. By contrast, the electron saturates the WGC by 19 orders of magnitude\footnote{While as remarked above it makes no sense to apply this constraint to massive gauge fields, we notice that the lightest fermions we know of, the neutrinos, have a mass close to \eq{cjsc}. This could perhaps be a sign they are coupled to a hidden massless $U(1)$ with not so small gauge coupling. Since the only anomaly-free $U(1)$ with generation-independent couplings in the SM with right-handed neutrinos is $B-L$, this would suggest the presence of extra particles charged under this $U(1)$ as well.}. 

Since in the SM the fermions receive mass from their coupling to the Higgs, \eq{sgc} also has implications for the cosmological hierarchy problem \cite{Giudice:2008bi,Csaki:2016kln,Giudice:2017pzm} (see also \cite{Ibanez:2017oqr,Craig:2019fdy} for some other proposed connections between the Swampland and the EW hierarchy problem). In particular, suppose one has a fermion with mass $m=yv/\sqrt{2}$, which in turn satisfies \eq{sgc}. Then 
\begin{equation} \frac{y^2}{2} v^2\gtrsim g M_P H= \frac{g}{\sqrt{3}} \rho^2,\end{equation}
where $\rho$ is the gravitating vacuum energy density and we have used $3M_P^2H^2=\rho^4$. Rearranging,
\begin{equation} \rho\lesssim \frac{y}{\sqrt{g}} v,\end{equation}
that is, no matter what the Higgs vev $v$ is, the gravitating vacuum energy (the one that actually enters in Einstein's equations) must be below the electroweak scale, by a factor of at least the smallest Yukawa of any charged fermion. There is a nice interplay with the arguments in \cite{Craig:2019fdy}, which introduced an additional gauge field to explain electroweak hierarchy. In this language, the electroweak hierarchy problem gets mapped to the question of why is the new gauge coupling small; \eq{sgc} could be part of the reason for this.

Equation \eq{sgc} does not lead to constraints on models of milli-charged dark matter, since the window excluded by the constraint has long been excluded experimentally \cite{Davidson:2000hf}.

One could also apply \eq{sgc} to $(B-L)$. This is an anomaly-free global symmetry of the SM Lagrangian, which is expected to be gauged or broken in quantum gravity. For instance, in GUT models, $(B-L)$ is often a spontaneously broken global symmetry, but it is not excluded that it could be massless, as long as the gauge coupling is small enough (around $g\lesssim10^{-24}$ \cite{Craig:2019fdy,Cheung:2014vva}). Our bound would mean that the lightest neutrino species cannot be massless.

Another topic we must address is inflation. During the inflationary phase the universe was approximately de Sitter, so it would be reasonable to apply \eq{sgc} there. Furthermore, as explained before, replacing the cosmological constant by a slowly-rolling scalar does not change our conclusions. We do not know much about the spectrum of massless $U(1)$'s and charged particles during inflation, but a naive expectation is that if inflation happens before EWSB, we have an $\mathfrak{su}(3)\oplus \mathfrak{su}(2)\oplus \mathfrak{u}(1)$'s worth of massless gauge fields to analyze\footnote{The same analysis would apply to GUT's, although inflation is usually taken to happen after GUT breaking to avoid a monopole problem \cite{Preskill:1979zi,Zeldovich:1978wj}.}. The charged spectrum of the SM remains massless at tree level (getting a mass of order Hubble scale via radiative corrections \cite{Chen:2016nrs}), which violates \eq{sgc}. We now discuss some possible ways out of this conundrum\footnote{We thank Prateek Agrawal for illuminating discussions.}:

\begin{itemize}

\item One can take inflation to only last $N_e\sim 60$ e-folds \cite{Baumann:2009ds}. Even if \eq{sgc} is violated, we could conceivably avoid the Big Crunch if the time that it takes for the black hole to discharge adiabatically is much higher than $N_e/H$. For a massless carrier, this time is set by the electric field only, $t_{\text{disch.}}\sim(qE)^{-1/2}$. Rearranging, this becomes a bound on the gauge coupling,
\begin{equation} g\leq \frac{H}{N_e^2M_P}.\label{gcbound}\end{equation}
Thus, we can avoid trouble, even with a large field inflation Hubble scale, by introducing a coupling between the inflaton and the gauge kinetic function of the gauge fields, in such a way that they remain small during inflation. This ensures that the gauge couplings become even more weakly coupled in the UV, while preserving gauge coupling unification as long as the coupling is taken to be same for every gauge field in the SM. This coupling would break any perturbative shift symmetry the inflaton field would have, so it is again fine-tuned unless it can be generated by non-perturbative effects. It would also lead to production of gauge fields during inflation and so, nongaussianity in the curvature or possibly curvature-magnetic field spectrum \cite{Maleknejad:2012fw,Jain:2012vm,Barnaby:2012tk}. These effects would be suppressed by at least a factor of \eq{gcbound} squared, so they would be small.

\item Another alternative is models of Higgs inflation \cite{Rubio:2018ogq},  where electroweak symmetry is still broken even at a high scale, and the fermions are massive. In vanilla models of Higgs inflation, the Higgs vev is comparable to $H^2 M_P^2$. This means that the fermions are still massive due to their Higgs coupling, but their masses are several orders of magnitude smaller due to the small Yukawa couplings, so \eq{sgc} is not satisfied. By contrast, in a more convoluted model with a complicated Higgs potential allowing for a hierarchy between fermion masses and the vacuum energy, this tension could disappear. 

In this scenario, one still has to think about $SU(3)$. The quarks are massive due to their coupling to the Higgs and thus pose no problem, but the gluons not commuting with the chosen $U(1)$ are charged, and naively massless. However, as long as the Hubble scale during inflation is below $\Lambda_{QCD}$, equation \eq{sgc} is satisfied in the same way as it is today. This corresponds to a vacuum energy density of $\sim10^{9}\, GeV$. Thus, a simple way out is a small field inflationary model. The price one pays for this is an extremely fine-tuned scalar potential, violating any reasonable form of the original no dS conjecture \cite{Obied:2018sgi,Agrawal:2018own}. It is worth noticing however that this is the same inflationary scale predicted by the recent TCC conjecture of \cite{Bedroya:2019snp} (the terminology ``transplanckian censorship'' was originally introduced in \cite{Draper:2019zbb} to refer to a different but possibly related phenomenon). 

Before wrapping up, we note that the non-abelian interaction terms in the presence of a background gauge field seem to give these gluons an extra contribution to the effective mass proportional to $\sqrt{g M_P H}$, when reduced to two dimensions to carry out the Schwinger calculation (see\cite{Ragsdale:2017wgi}). Still, this is at best just marginally satisfying \eq{sgc}, so without further analysis it is unclear whether the crunch is really avoided. Having $\Lambda_{QCD}< H$ is a sure way to avoid the trouble caused by the gluons.
\end{itemize}

\subsection{Weak gravity in de Sitter space}\label{sec:wgc}

We have a constraint on the EFT from black hole decay, but it would be nice if we could recover WGC as well. There are immediate obstacles to this. The usual motivation for the (mild form of the) WGC in flat space or AdS is that otherwise all extremal black holes are exactly stable or marginally unstable (as is the case for BPS black holes in a theory with BPS particles), and this is somehow undesirable. In flat space it is still unclear what precisely goes wrong if this does not happen, but in AdS a violation of the mild form of the WGC conflicts with properties of entanglement entropy in the dual CFT \cite{Montero:2018fns}. 

By contrast, we have seen in the previous Section that, the moment there is any charged particle in the spectrum, RN-dS black holes follow the fate of their uncharged flat space counterparts -- they eventually evaporate completely--. This is easy to understand intuitively. In flat space or AdS , a subextremal charged particle that tunnels out of a black hole has no choice but to fall back in, because it just does not have enough energy to escape to infinity\footnote{In the AdS case the analysis is more subtle, since even massive superextremal particles cannot escape to infinity. Instead, they form a charged cloud which completely discharges the black hole in the extremal case \cite{Denef:2009tp}.}. In dS however, even a very massive particle will sometimes tunnel so far away from the black hole that it is pulled away by the cosmological expansion, being eventually eaten up by the cosmological horizon. 

In fact, black holes discharge even in the absence of charged particles. In that case, they move towards the lukewarm line in Figure \ref{fig1}, or towards the charged Nariai branch if the initial charge happens to be higher than the intersection point of the charged Nariai branch with the lukewarm line. Both on the lukewarm line and on the charged Nariai branch there are gravitational instantons \cite{Ginsparg:1982rs,Bousso:2002fq} that mediate transitions from these black holes to and from empty de Sitter space. But the point of our analysis is that, even if one demands that extremal black holes must be unstable, this leads to no constraints on the spectrum of Einstein quantum gravity in dS space (or situations close to it, like slow-roll quintessence), since black holes decay automatically anyway. This was already foreseen in \cite{Dias:2018etb}.

Although we have just argued that no charged black holes are stable in dS, small black holes have very large lifetimes if WGC is not satisfied. Charged particles must traverse an incredibly wide potential barrier, from the black hole horizon to the cosmological one, if the black hole is to discharge.  The current associated to this effect is therefore exponentially exponentially suppressed in $\ell$; one can see this in the exact expression for the Nariai current \eq{KKmodes2}. 

To sum up, in a dS-like theory which contains no WGC particles, small black holes do evaporate, but their lifetime is exponentially long in $\ell$. There are so long-lived that it makes sense to include them in the computation of the Schwinger current for a large Nariai black hole. As mentioned in Subsection \ref{sec:sgc}, there are two ways to trigger the adiabatic regime. One way is to have one or a few particles violating \eq{sgc}. Another is to have so many particles satisfying \eq{sgc} that their combined current beats the Schwinger exponential suppression, resulting in an unsuppressed current. The contribution from black holes to the current is\footnote{We have only included extremal black holes and replaced the sum over integer-quantized charges $Q_r$ by an integral. We have neglected the $S$ term and then taken the upper integration limit to be $\infty$ since the integral is dominated by small black holes as we take $Q_{\text{max.}}$ much smaller than a cosmological size black hole. For the same reason, we have also used the flat-space extremality expression. We have taken $E=g M_P H$ to be a typical electric field in the upper part of the Nariai branch.}
\begin{equation} \mathscr{J}_r\approx\int_{\frac{\Lambda}{\sqrt{2} g M_P}}^{Q_{\text{max.}}}dQ_r\, e^{S-\frac{M^2}{Q_rE}}\sim \left(\frac{H}{M_P}\right)\frac{H^2\Lambda^2}{g} e^{-4\sqrt{2}\pi \frac{\Lambda}{H}} \label{gvs}\end{equation}
Here, $\Lambda$ is some EFT cutoff, which cuts off the sum when black holes become so small than the semiclassical description breaks down. $S$ is the black hole entropy counting the number of black hole states with the given charge and mass. The upper bound on the charge $Q_{\text{max.}}$ is there because here we just want to consider small black holes -- the entropy contribution is then subleading and the result is dominated by the lower integration limit. Equation \eq{gvs} grows without bound as $g\rightarrow0$ which means that, as long as this limit is allowed (and it might very well not be; this is just a heuristic argument), we will enter the adiabatic regime. Thus, we obtain an argument for ordinary WGC in de Sitter, that small black holes should be allowed to decay, with a similar flavor to the original heuristics in \cite{ArkaniHamed:2006dz}. One might also conclude that $g\rightarrow0$ is not allowed; this is in fact the conclusion when one follows the same argument in AdS \cite{Montero:2017mdq}. 

This argument is heuristic and it relies on a very unclear limit. We also tried to do better, and in the remainder of the Section we report what we got. Since WGC guarantees that black holes evaporate quickly enough, it is reasonable to imagine it is related to (an upper bound on) evaporation times of black holes. Interestingly, one can argue something like this -- an upper bound on the \emph{equilibration} time of a black hole--. The equilibration time $t^{\epsilon}_{\text{eq.}}$ of a particular quantum state is the time that it takes for the state to become close to the thermal state to any desired accuracy $\epsilon$. 

In our context, and as mentioned above, we should regard black hole evaporation as (part of) an equilibration process towards the thermal state, represented by empty dS. Once a black hole evaporates, the decay products fly away and eventually fall behind the cosmological horizon. When this happens, to any desired degree of accuracy, one can say the state has finally equilibrated. 

We refer the reader to the excellent review \cite{Gogolin:2016hwy} for a precise definition and more details. Interestingly, it is possible to derive a universal upper bound on the equilibration time of a generic quantum state\footnote{Assuming it thermalizes eventually.} (see Theorem 1 of \cite{Gogolin:2016hwy}), which grows exponentially with the system size\footnote{This kind of scaling is achieved by relaxation times in glassy systems.}. Notice that this is still better than the quantum Poincar\'{e} recurrence time $e^{e^{S_{dS}}}$. Thus, we might want to demand that the total evaporation time of a black hole is faster than $e^{S_{dS}}\ell$. The evaporation time for a charged black hole is controlled by the exponential factor of the Schwinger current, $e^{-m^2/(qE)}$. So a very naive realization of this bound could be to demand that $m^2/(qE)<S_{dS}$, which leads to a lower bound on the charge-to-mass ratio of the particle that mediates black hole decay,
 
\begin{equation}\frac{g q M_P}{m}\gtrsim \frac{ H m}{M_P^2}.\end{equation}

While this bound goes on the same direction as the WGC, it is parametrically weaker and so uninteresting. It becomes trivial as $H\rightarrow0$, so it is clearly not capturing the actual physics of WGC. We just want to illustrate that upper bounds on equlilibration times lead to lower bounds on the charge-to-mass ratio; and while the universal result in \cite{Gogolin:2016hwy} does not lead to an interesting result, a stronger lower bound derived using the particular properties of the system under study might be behind the rationale for WGC in quasi-de Sitter spacetimes.

\section{Conclusions}\label{sec:conclus}

In flat space and AdS, the Weak Gravity Conjecture is intimately related to Cosmic Censorship and the stability of extremal black holes. In this note, we have tried to understand to what extent this kind of reasoning works in de Sitter, and more generally to spacetimes with positive vacuum energy, and to figure out if there are any new candidates for Swampland constraints. 

To do so, first one needs to understand how black holes evaporate in de Sitter space. Unlike their flat space counterparts, whose mass for a given charge is bounded below by the extremality curve but unbounded from above, de Sitter black holes have a largest mass for a given value of $Q$ -- the so-called Nariai solution. This is a genuinely de Sitter effect, and one can connect it to the idea that the Hilbert space in the static patch should be finite-dimensional \cite{Banks:2000fe,Banks:2003cg,Banks:2005bm,Witten:2001kn}, from which it follows that there is a maximum value of any reasonably well-defined observable like the mass or the charge.  

Previous studies of black hole evaporation in de Sitter \cite{Bousso:1996au,Bousso:1997wi,Bousso:1999ms,Niemeyer:2000nq,Belgiorno:2008mx,Belgiorno:2009da,Belgiorno:2009pq,Kim:2015wda,Belgiorno:2016nup} have not examined how the mass and charge of a black hole evolves in time in a theory with charged particles. We have worked out how large de Sitter black holes evaporate, depending on the spectrum of charged particles in the theory. For a single scalar particle of mass $m$ and charge $q$, we find two very different regimes, controlled by the parameter $\chi\equiv m^2/(gq M_P H)$:\begin{itemize}
\item When $\chi\gg1$, the Schwinger pair production that allows black holes to lose charge is exponentially suppressed. They therefore evaporate slowly, schematically following the curves illustrated in Figure \ref{shark-flow}. All black holes evaporate eventually, never leaving the ``shark fin'' region of nice sub-extremal solutions. This is consistent with the thermodynamic interpretation of de Sitter as a thermal system at finite temperature \cite{Banks:2005bm,Anninos:2012qw}. According to this idea, perturbations in de Sitter should eventually evolve back to empty de Sitter space, and this process corresponds to equilibration with the thermal bath. Field theory modes and small black holes all agree with this picture, and it was known that large neutral black holes do so as well \cite{Bousso:1997wi}. We have shown that this is the case for charged black holes as well, as long as $\chi\gg1$.
\item If $\chi\ll1$ instead, a charged Nariai black hole sees its electric field immediately screened out by the very light charged carriers produced via Schwinger; the electric field is replaced by radiation, which cannot support the solution and collapses to a Big Crunch. This Big Crunch is of the same kind that the one which appears in the future of a superextremal, neutral Schwarzschild-de Sitter solution with mass above the Nariai value. Morally, the black hole loses all the charge instantaneously, while keeping its mass constant, as illustrated in Figure \ref{fig6ymedio}.
\end{itemize}
The first noteworthy (if unsurprising) consequence of these results is that, as advanced in \cite{Dias:2018etb}, there is no kinematic barrier to black hole decay. This is unlike in flat space, where at least one particle must satisfy WGC if extremal black holes are to decay. In flat space, a subextremal charged particle can tunnel out of an extremal black hole, but it inevitably falls back because for it gravity is not the weakest force. By contrast, the same particle in dS can always tunnel far away enough that the cosmological force overcomes the gravitational attraction from the black hole. Gravity may not be the weakest force, but in any case it is always weaker than the cosmological one. 

This is however not the end of the story. In the second scenario, where the black hole discharges quickly, there is a failure of the thermalization picture; there are black holes that become superextremal, crunch, and never thermalize back to de Sitter space. We lack a complete, UV description of Einstein quantum gravity in long-lived de Sitter space (if it exists at all), so we do not know if this is really an inconsistency. However, we cannot help to point out that the superextremal region in Figure \ref{fig1} is connected, so one can continuously deform any superextremal solution into any other. A neutral superextremal black hole can be connected to a small superextremal one, which we ordinarily like to avoid based on cosmic censorship or more sophisticated ideas (like the fact that these solutions do not have any known embedding in string theory and it would be difficult to give a thermodynamic interpretation). 

In any case, if one wants the thermal picture of the dS static patch to hold, then our results suggest that one should impose $\chi\gtrsim1$ as well. This becomes a constraint on the effective field theory, that
\begin{equation*} m^2\gtrsim q g M_P H,\end{equation*}
for every particle in the theory. This is a constraint on the effective field theory coming from gravitational arguments, and since we do not know of any field theory arguments that would lead to it, it has the flavor of a Swampland constraint. However, since we cannot check against stringy examples and it relies on the thermal picture of the dS static patch and its finite-dimensional Hilbert space, it is on shakier ground than actual Swampland constraints. It is just a consistency condition we need to enforce so that the thermal picture of dS can hold at all. 

Interestingly, avoiding the adiabatic regime also leads to a new heuristic argument for the WGC from black hole decay in dS, circumventing the aforementioned lack of kinematical constraints. In a WGC-violating theory, small extremal black holes would be very long-lived, and would contribute to the discharge of Nariai ones. In the limit of small gauge coupling, there is so many of them that they can trigger the onset of the adiabatic regime by themselves. A natural way to avoid this is then the WGC. 

This argument is very similar to the original WGC heuristic, and suffers from the same weaknesses. For instance, it only works in the limit $g\rightarrow0$, and we have no idea whether this limit makes any sense in any way. Trying to address this, we have speculated that WGC might be related to an upper bound in evaporation times of black holes. In the thermal picture of dS, this would correspond to an upper bound to equilibration times; while there are universal results in this direction which lead to WGC-like bounds, they are much weaker. 

Since the real world looks like de Sitter at least for now, it is reasonable to compare this constraint with it. Taking the $U(1)$ to be electromagnetism, the constraint turns out to be satisfied by every charged particle we know of. Again, this is a UV constraint, since there is nothing obviously wrong from the EFT point of view with a world where the electron was lighter than neutrinos or the cosmological constant was a few orders of magnitude higher\footnote{We wouldn't exist, but it is unclear that this signals an actual inconsistency of the theory.}.

Since the constraint ties charged fermion masses to the Hubble scale, it alleviates to some extent the cosmological hierarchy problem. It also constrains models of mili-charged dark matter, but the window where the constraint is nontrivial has been long excluded by experimental data.

The constraint has an interesting interplay with inflation. Since during inflation the universe was basically de Sitter for 60 efolds, our arguments apply, not just to electromagnetism but to whichever massless unconfined gauge fields are present during the inflationary era. This puts constraints on inflationary models. We have sketched a couple ways to comply with these constraints: Direct inflaton-gauge kinetic term couplings (which would lead to small nongaussianities), or models of Higgs inflation with a plateau potential and an inflationary scale below $\Lambda_{QCD}$. 

A number of interesting research directions remain. Tracing out the boundary of Figure \ref{fig1} by starting with an infinitesimally small extremal black hole, moving up along the extremal branch up to the ultracold point and then moving back down along the charged Nariai branch, one finds a remarkable situation. There is a continuous interpolation between $AdS_2 \times S^2$, $\text{Mink}_2 \times S^2$, and $dS_2 \times S^2$. It would interesting to understand this as a continuum of two-dimensional theories, especially in the context of the recent surge of interest in two-dimensional quantum gravity \cite{Jensen:2016pah,Maldacena:2016upp,Cotler:2019nbi,Maldacena:2019cbz,Anninos:2019oka,Saad:2019lba}.

A natural extension of our work would be to higher dimensions. One could study both higher dimensional de Sitter spaces and black branes rather than black holes. It would be interesting to see if this produces bounds on brane tensions analogous to the bounds on particle masses obtained here. Since in the Nariai branch we get a $dS_2\times S^2$ geometry for black holes, it is natural to guess that by looking at black branes one could get higher-dimensional de Sitter solutions supported by flux which discharge slowly, and it would possible to study their decay in a UV-regulated way. 

\section*{Acknowledgments}
We thank Oscar Dias, Ben Freivogel, Gui Pimentel, Jorge Santos, Raphael Bousso, Jaume Garriga, Dionysios Anninos, Aron Jansen, Matt Reece, Ben Heidenreich, Tom Banks, Grant Remmen, Isabel Garc\'{i}a-Garc\'{i}a, J. L. F. Barb\'{o}n, Thomas Hertog, John Stout, Jan Pieter van der Schaar, and especially Prateek Agrawal and Patrick Draper for very useful discussions and comments. 

The  work  of  VV  is  supported  by  the  European  Research  Council grant  ERC-2013-CoG  616732  HoloQosmos.  The work of TVR and VV is supported by the FWO Odysseus grant G.0.E52.14 and the C16/16/005 grant of the KULeuven. MM was supported by an FWO Fellowship throughout the completion of this project except in the final stages, where he is supported by grant 602883 from the Simons Foundation. We thank IFT Madrid and the organizers of the Workshop ``Navigating the Swampland'', the 2019  String Theory and Cosmology Gordon Research Conference and Seminar, and the Aspen Center for Theoretical Physics under the ``String Theory and the Hidden Universe'' program for creating a stimulating environment in which parts of this work were carried out.

\appendix
\section{Details about the quasistatic equations of motion}\label{app:EEex}
Here we provide some details about the semiclassical equations of motion \eq{1eoms} quoted in the main text. To first order in $\epsilon$, the coefficients of the Einstein tensor expansion are
\begin{align}\delta_1&= \frac{1}{2} \left(\frac{2 {\delta A} \left(3 M^2 r^2-5 M Q^2 r-3 M r^5+M r^3+2 Q^4+\left(2 Q^2-1\right) r^4+2 r^6\right)}{r^4 \left(2 M r-Q^2+r^4-r^2\right)}\right.\nonumber\\&\left.+\frac{2 {\delta B} \left(M^2 r^2+M Q^2 r+7 M r^5-M r^3-Q^4-4 Q^2 r^4+r^8-2 r^6\right)}{\left(2 M r-Q^2+r^4-r^2\right)^3}\right.\nonumber\\&\left.+\frac{r^2 {\delta B}''}{-2 M r+Q^2-r^4+r^2}+\frac{r \left(-3 M r+2 Q^2+r^2\right) {\delta B}'}{\left(-2 M r+Q^2-r^4+r^2\right)^2}-\frac{8 Q {\delta F}}{-2 M r+Q^2-r^4+r^2}\right.\nonumber\\&\left.+\frac{\left(-3 M r+2 Q^2+r^2\right) {\delta A}'}{r^3}\right),\\\delta_2&=\frac{1}{2} \left(-\frac{2 {\delta A}  \left(r \left(-M^2-M r^3+3 M r+2 r^6-r^2\right)+Q^2 \left(M+2 r^3-2 r\right)\right)}{r^3 \left(2 M r-Q^2+r^4-r^2\right)^{-1}}\right.\nonumber\\&\left.+\frac{2 {\delta B} \left(r^2 \left(-5 M^2-5 M r^3+3 M r+r^6\right)+Q^2 r \left(5 M+4 r^3-2 r\right)-Q^4\right)}{2 M r-Q^2+r^4-r^2}\right.\nonumber\\&\left.+r^2 \left(2 M r-Q^2+r^4-r^2\right) {\delta B}''+8 Q {\delta F} \left(-2 M r+Q^2-r^4+r^2\right)\right.\nonumber\\&\left.-\frac{\left(M+2 r^3-r\right) \left(-2 M r+Q^2-r^4+r^2\right)^2 {\delta A}'}{r^2}+r^2 \left(-M-2 r^3+r\right) {\delta B}'\right),\end{align}\begin{align}\delta_3&=\frac{4 r \left(r \dot{M}-Q \dot{Q}\right)}{-2 M r+Q^2-r^4+r^2},\\\delta_4&=\frac{1}{2} \left(-\frac{2 {\delta A} \left(r \left(-M^2+5 M r^3+M r+5 r^6-4 r^4\right)-Q^2 \left(-M+r^3+r\right)\right)}{r^5}\right.\nonumber\\&\left.-\frac{2 {\delta B} \left(r \left(M^2+7 M r^3-M r+r^6-2 r^4\right)+Q^2 \left(-M-5 r^3+r\right)\right)}{r \left(-2 M r+Q^2-r^4+r^2\right)^2}\right.\nonumber\\&\left.-\frac{\left(-3 M r+2 Q^2+r^2\right) {\delta B}'}{r \left(2 M r-Q^2+r^4-r^2\right)}-\frac{\left(M+2 r^3-r\right) \left(2 M r-Q^2+r^4-r^2\right) {\delta A}'}{r^4}-\frac{4 Q {\delta F}}{r^2}\right.\nonumber\\&\left.+{\delta B}''\right). 
\end{align}
To first order in $\epsilon$, Einstein's equations are simply $\delta_i=8\pi G \eta_i$. The $\delta_3$ component gives \eq{1eoms} in the main text; here we focus on the remaining components, that give the equations of motion for $\delta A$, $\delta B$, $\delta F$.  We have omitted the dependence on the slow time scale $t_1$, since there are no time-derivatives anywhere in these expressions. The perturbation $\delta F$ also shows up in Maxwell's equations, but these are redundant, as Einstein's equations imply energy-momentum conservation $\nabla_a T^{ab}=0$ which is in turn equivalent to Maxwell's equations (this is the case because we only have one field; in the quintessence discussion in Appendix \ref{app:q}, one needs to introduce an additional function and impose the equation of motion for the radial perturbation by hand, as we do in the main text). 

Since there are no time derivatives, the time dependence on the $\delta_i$ is fixed by the time dependence of the $\eta_i$.  For the cases under consideration, where $\eta_i$ comes from Hawking or Swchinger effects, these only depend on $t_1$, since the stress-energy for these depend on time only through $M$ and $Q$. 

To sum up, we have a system of three coupled second-order ODE's for the perturbations with non-singular coefficients outside the horizons. Working in first-order formalism, $\delta B''(r)=C'(r)$, the equations $\delta_i=8\pi G \eta_i$ together with $C=\delta B'(r)$ can be written as a matrix ODE
\begin{equation} \mathbf{M}_1\vec{y}'+\mathbf{M}_2 \vec{y}= \vec{\eta},\end{equation}
where $\vec{y}=(\delta A, \delta B, \delta F, C)$, $\vec{\eta}=8\pi G (\eta_1, \eta_2, \eta_4,0)$, and $\mathbf{M}_1,\mathbf{M}_2$ are coefficient matrices. Crucially, 
\begin{equation}\text{det}(\mathbf{M}_1)=\frac{Q \left(r \left(-2 M-r^3+r\right)+Q^2\right)^2}{r^3}\neq0,\end{equation}
so $\mathbf{M}_1$ is invertible. Then, as long as there are no singularities in the $\eta_i$ between the two horizons, existence of the solution given boundary conditions is then guaranteed by the general theory of ODE's e.g. Caratheodory's theorem \cite{rudin1987real}. 

Since we have solved all of Einstein's equations, our solutions satisfy the Hamiltonian and momentum constraints, to first order in $\epsilon$. Again, we can generically expect violation of the Hamiltonian constraints to take place at second order in $\epsilon$.

\section{Generalization to quintessence}\label{app:q}
In this Appendix, we describe the generalization of the quasistatic and adiabatic equations of motion \eq{1eoms} when one introduces a very slowly rolling scalar field $\varphi$. This will be relevant for us because the conjecture in \cite{Obied:2018sgi,Agrawal:2018own} suggests that exactly stable Einsteinian de Sitter solutions do not exist; hence, we need to know to what extent are our results robust against a small slow-roll.

\subsection{Quasistatic}\label{app:qquasi}
We will parametrize the rolling of $\varphi$ by a time-dependent Hubble radius $\ell(t)$, on top of $M(t)$ and $Q(t)$. So in this Appendix we will restore $\ell$. The scalar will couple via a Lagrangian
\begin{equation}\mathcal{L}_\varphi=\frac{1}{2\epsilon} (\partial\varphi)^2-V(\varphi),\label{scalpot}\end{equation}
where $\epsilon$ is a small parameter. The method of separation of scales requires an unperturbed problem and a parametric family of solutions for it. We will choose the unperturbed problem to be the $\epsilon\rightarrow0$ limit of \eq{scalpot}. The family of solutions are the RN-dS solutions with the three parameters $(Q,M,\ell(\varphi))$, where the Hubble radius is a function of the scalar potential given by
\begin{equation} \frac{3}{\ell(\varphi)^2}=V(\varphi_0).\label{qdsf}\end{equation}
To fully specify the family of solutions, we also need to specify the profile for the quintessence field. Equation \eq{qdsf} implies that it is a constant. This is consistent with the equations of motion of the system, which are
\begin{equation}G_{\mu\nu}=8\pi G (T^{\text{EM}}_{\mu\nu}+ V(\varphi)g_\mu\nu),\quad \nabla_\mu\nabla^\mu \varphi=0.\end{equation}
Now, we work to first order in $\epsilon$, and introduce a small time dependence,
\begin{equation} M(\epsilon t),\,Q(\epsilon t),\,\ell(\epsilon t)=\frac{3}{V(\varphi_0(\epsilon t))}.\end{equation}
On top of this, we introduce order $\epsilon$ but time-independent perturbations to the metric, gauge field and quintessence. The metric and gauge field components work out in the same way as above, so in order not to clutter the equations we will not re-discuss them. For quintessence, we have
\begin{equation}\varphi(r,t)=\varphi_0(\epsilon t) + \epsilon\, u(r),\end{equation}
with an equation of motion $\nabla_\mu\nabla^\mu \varphi=\epsilon V'(\varphi)$. To first order in $\epsilon$, we get
\begin{equation}g^{tt} \ddot{\varphi}+ \frac{1}{r^2}\partial_r\left( g^{rr} r^2 u'(r)\right)=V'(\varphi_0(0)),\end{equation}
which fixes $\ddot{\varphi}=0$ (familiar from slow roll approximation) and ODE which uniquely determines $u(r)$, given boundary conditions. Einstein's equations become as before give equations of motion for the time-independent metric perturbations, and the off-diagonal components as before give the equation of motion for $M$\footnote{Notice that the equations are written in terms of $\mathcal{T}$, $\mathcal{J}$ instead of $\mathscr{T}$, $\mathscr{J}$.},
\begin{equation}\frac{dM}{dt}=-\pi G \ell^2 r^4  U\mathcal{T}-\frac{4 \pi  \mathcal{J}}{Q}+r^2\frac{{\dot{\ell}}}{\ell ^3}{r}.\label{qeveom}\end{equation}
The extra last term is the contribution from quintessence. The equation of motion for $\dot{Q}$ remains the same. Out of the $\delta_4$ component of the Einsteins equations we also get
\begin{equation}-\frac{6\dot{l}}{\ell^3}=V'(\varphi_0(0)) \dot{\varphi}_0(0),\end{equation}
which is consistent with \eq{qdsf}. 

So now that we have the modified equations of motion, we just need to check what motion do they enforce on the $(M,Q,\ell)$ space. To check whether we leave the Nariai branch, again, we just need to compute $\dot{\Delta}$, where $\Delta$ is the discriminant of the quadratic equation $U(r)=0$. We cannot neglect the $\ell$ dependence in the discriminant now,
\begin{equation}\Delta = \ell^4 M^2-\ell^4 Q^2-27 \ell^2 M^4+36 \ell^2 M^2 Q^2-8 \ell^2 Q^4-16 Q^6,\end{equation}
and becoming superextremal is equivalent to $\dot{\Delta}>0$ on the Nariai branch. Using \eq{qeveom}, one gets \begin{equation} \dot{\Delta}=0,\end{equation}
so that again one moves along the charged Nariai branch, without changing the charge. This is precisely what \cite{Bousso:1996au} found for the extremal case. Their analysis is in fact general for classical effects on the charged Nariai branch, and suggests that other modifications such as e.g. introducing a coupling of $\varphi$ to the gauge field cannot change the picture; we have checked explicitly that this is indeed the case. 

\subsection{Adiabatic}\label{app:qadiab}
We will only analyze the Nariai branch, as in the main text. The 2d effective action is now 
\begin{equation}8\pi G\, S=\int d\Omega_2 d\tau dx\left\{\dot{(e^{2\phi})} \dot{a} -e^{-\phi}a+\left(V(\varphi)-\frac12\dot{\varphi}^2\right)e^{\phi} a\right\}+8\pi G \int dV_2d\Omega_2 \sqrt{-g}\mathcal{L},\label{act0q}\end{equation}
where the quintessence potential and kinetic term replace the $+3$ cosmological constant term in \eq{act0q}. We will assume no coupling to the electromagnetic field. Defining again $s\equiv e^{2\phi}$  and assuming isotropic radiation stress-energy, as in the main text, the equations of motion are 
\begin{align}\ddot{\varphi}&=-\frac{V'(\varphi)}{\sqrt{s}}-\left[\frac{\dot{a}}{a}+\frac{\dot{s}}{s}\right]\dot{\varphi},\label{eom9}\\
\ddot{s}&=V(\varphi)\sqrt{s}-s\dot{\varphi}^2-\frac{1}{\sqrt{s}}\left(1+\frac{\rho_0}{a^{4/3}}\right),\label{eom10}\\ \frac{\ddot{a}}{a}&=\frac{1}{2s^{3/2}}\left(1-\frac{\rho_0}{a^{4/3}}+V(\varphi)\,s\right)-\dot{\varphi}^2,\label{eom11}\end{align}
where $\rho_0$ is again the energy density of the electromagnetic field just before decay, and we have assumed that $V(\varphi_0)=3$, so that we can use expressions like \eq{Ur}. 

\begin{figure}
\begin{center}
\includegraphics[width=0.55\textwidth]{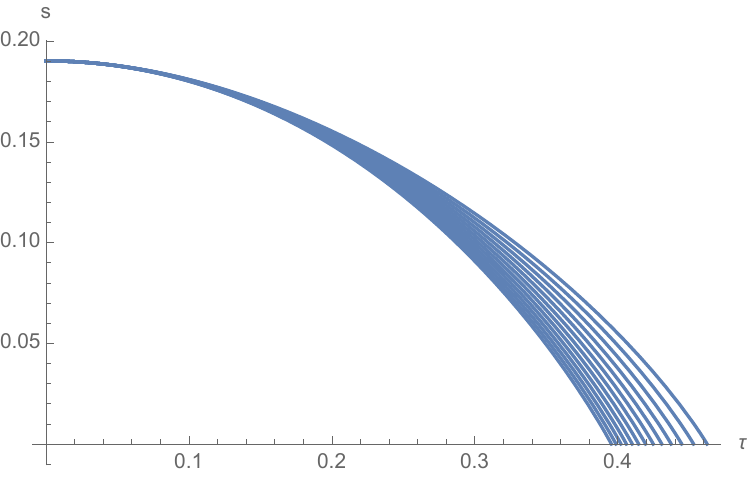}
\vspace{0.5cm}

\caption{\small{Numerical solution for $s(t)$ of the system of equations \eq{eom9}-\eq{eom11} for an initial charge $Q=0.99 \cdot12^{-1/2}$ and different values of $\alpha$ from $-1/3$ to $+1/3$ in $1/6$ increments. Central value corresponds to $\alpha=0$. This illustrates that reasonable quintessence potentials are not enough to avoid the Big Crunch on the charged Nariai branch.}}\label{quintspread}
\end{center}
\end{figure}

The question is whether the addition of quintessence can alleviate or even make the crunch disappear. We will limit ourselves to a numerical analysis illustrating these qualitative features in the particular case of an exponential potential $V=3e^{-\alpha\varphi}$. Results are presented in Figure \ref{quintspread}. Quintessence causes a slight delay of the crunch for negative values of $\alpha$, but this effect is small unless $\vert\alpha\vert\gg1$, in which case the model does not make sense anyway.

\section{Exact Schwinger pair production on the charged Nariai branch}\label{app:garr}
Here we compute the exact evolution equation on the Nariai branch, using the results in \cite{Frob:2014zka}. The steady current for a particle of two dimensional mass $m_{2d}$ and charge $q$ is 
\begin{equation} \mathscr{J}_{dS_2,r}(m_{2d},q)=\frac{qH_{dS_2}}{\pi}\frac{\sigma}{\sinh(2\pi \sigma)} \sinh(2\pi qE/H_{dS_2}^2),\label{jds2}\end{equation}
where\footnote{Under dimensional reduction we have that $1/{g_{2d}^2}=\frac{4\pi r^2}{g_{4d}^2}$ and $E_{4d} =\frac{g^2_{4d}Q_r}{4\pi r^2}=g_{2d}^2Q_r q=E_{2d}$, so both the product $qE$ and $m$ can be computed in 4d and then one can use the 2d formula directly. The normalization factor in the scalar accounts for integration over the sphere.}
\begin{equation}\sigma=\sqrt{\frac14-\frac{m^2}{H^2}-\frac{q^2E^2}{H^4}},\quad qE= \sqrt{\frac{1}{4\pi G}}\frac{g q}{\ell} \frac{Q\ell^2}{r_g^2},\label{efn}\end{equation}
and
\begin{equation}r_g^2=\frac{\ell^2}{6}\left(1+\sqrt{1-12Q^2}\right),\quad H^2_{dS_2}\ell`2=6\left(\frac{1}{1+\sqrt{1-12Q^2}}-1\right).\end{equation}
It is convenient to introduce parameters
\begin{equation}\chi\equiv \frac{qE}{\tilde{m}^2}=\frac{qE}{m^2-H^2/4}\end{equation}
and
\begin{equation} y\equiv \tilde{m}^2/H^2.\end{equation}
In terms of these, equation \eq{jds2} becomes
\begin{equation}\mathscr{J}_{dS_2,r}(m_{2d},q)=\frac{q}{\pi} H_{dS_2}\frac{\sqrt{y^2\chi^2+y}}{\sinh(2\pi \sqrt{y^2\chi^2+y})}\sinh(2\pi \chi y).\end{equation}

Notice that
\begin{equation}\chi y= \frac{qE}{H_{dS_2}^2}>0.\end{equation}
If one takes $y>0$, which amounts to having a field more massive than $H_{dS_2}^2/4$, it is possible to write down a simple lower bound on the current. For any $a>0,b>0$ we have
\begin{equation}\frac{\sinh(a)}{\sinh(\sqrt{a^2+b})} \geq  \frac{\tanh(a)}{e^c},\end{equation}
where $c\equiv \sqrt{a^2+b}-a\leq \frac{b}{2a}$. Due to this, one can write
\begin{equation} \frac{\sinh(a)}{\sinh(\sqrt{a^2+b})} \geq \tanh(a)e^{-\frac{b}{2a}}.\end{equation}
Now taking $a=2\pi\chi y$, $b=4\pi^2 y$, one can bound
\begin{equation}\mathscr{J}_{dS_2}(m_{2d},q)\gtrsim\frac{q}{\pi} H_{dS_2}\frac{\sqrt{y^2\chi^2+y}}{\sinh(2\pi \sqrt{y^2\chi^2+y})}\sinh(2\pi \chi y)\geq \frac{q^2E}{\pi H_{dS_2}}\tanh\left(2\pi\frac{qE}{H_{dS_2}^2}\right)e^{-\frac{\pi}{\chi}}.\label{loki}\end{equation}

The bound becomes tighter and tighter the higher the hierarchy between $m^2$ and $qE$. The factor $e^{-\frac{\pi}{\chi}}$ is precisely the Schwinger space prefactor \eq{schw00}; the bound becomes tight when $y\gg1$, which means that the curvature effects become unimportant and the pair-production happens essentially in flat space.

As in the main text, a 4d field of mass $m$ and charge $q$ decomposes into a tower of 2d fields with masses
\begin{equation} m_{2d}(s)^2=m^2+\frac{s(s+1)}{r_g^2}.\end{equation}
Here, $s=0,1,2,...$ and there is a multiplicity $2s+1$ for the $s$th mass. Note that depending on the spectrum of the four-dimensional theory, it may well be possible that $m_{4d}^2 \gg \frac{s(s+1)}{r_c^2}$ and the KK modes are densely packed compared to the lowest mass. It is then important to sum over the KK modes when computing the Schwinger current. This is reflecting that the physics of the problem is intrinsically four-dimensional, but simple enough that it can be analyzed in 2d terms. 

Then the current equation of motion in \eq{eom1} becomes a sum over KK modes, which we can  evaluate explicitly using \eq{loki}, since only the last factor depends on $m$. In this way, \eq{loki} becomes
\begin{equation}\mathscr{J}_r\gtrsim\frac{q^2r_c^2}{\pi^2} \vert E\vert  H_{dS_2}\frac{\sqrt{y^2\chi^2+y}}{\sinh(2\pi \sqrt{y^2\chi^2+y})}\sinh(2\pi \chi y)\geq \frac{q^2r_c^2E\vert E\vert }{\pi^2 H_{dS_2}}\tanh\left(2\pi\frac{qE}{H_{dS_2}^2}\right)e^{-\frac{\pi}{\chi}}.\label{loki2}\end{equation}
The electric field on the Nariai branch is given by \eq{efn}. Using this, equation \eq{eom1} then becomes\footnote{The prefactor accounts for the definition of $Q$ \eq{adimp} in terms of the actual integer-valued current. The factor of $4\pi r_g^2$ in \eq{2eoms} is already taken into account in the expression for $\mathscr{J}_{dS_2}$, since this is a two-dimensional calculation.}
\begin{equation}\mathscr{J}=\sqrt{\frac{g^2G}{4\pi \ell^2}}\mathscr{J}_r\gtrsim\sqrt{\frac{g^2G}{4\pi \ell^2}}\frac{q^2r_c^2E\vert E\vert }{\pi^2 H_{dS_2}}\tanh\left(2\pi\frac{qE}{H_{dS_2}^2}\right)e^{-\frac{\pi}{\chi}}\label{KKmodes1}\end{equation}
This is equation \eq{KKmodes2} of the main text.

\section{Big Crunch and singularity theorems}\label{app:crunch}
In this Appendix we show that the Big Crunch in the adiabatic approximation discussed in the main text is actually independent of the details of the model and is in fact guaranteed by a singularity theorem. The starting point of our discussion will be the dimensionally reduced Lagrangian to two dimensions of the Einstein-Maxwell system,  \eq{metricansatz} of the main text. However, we will now have a general two-dimensional metric $g_2$, and in the two-dimensional Lagrangian $\mathcal{L}$ we will include a sum over all the KK modes of metric components and any matter fields we include. By working high up the Nariai branch, we can have a parametric hierarchy between the size of the two-spheres on which we reduce the theory and the non-compact space, so we mostly are thinking of the regime where a honest KK reduction is possible, but our discussion is general.

We will apply one of the original singularity theorems \cite{Hawking:1973uf}. In the main text, we picked a particular spacelike hypersurface in the Nariai geometry, declared that this was the ``starting point'' of the evolution, and then let the electric field be screened out. We want to prove a crunch exists too for an arbitrary initial data hypersurface $\mathcal{C}$, even one that crosses the cosmological or black hole horizons. We also want to work (slightly) outside of the Nariai limit. Thus, we need to consider the maximal analytic extension of the RN-dS black hole, a spacetime described in \cite{Bousso:1998bn}. Removing singularities, it has topology $S^2\times S^1\times \mathbb{R}$. In particular, any global Cauchy slice $\mathcal{C}$ is compact. The singularity theorems have a loophole precisely in this case \cite{Senovilla:2014gza}. One can lift this problem however by working in the universal cover of the RN-dS black hole, which describes an infinite sequence of ``Bousso beads''. This is the spacetime we will work in. 

For a particular Cauchy surface $\mathcal{C}$, the singularity theorem guarantees that every geodesic is not future-extendible provided that \begin{enumerate}
\item The expansion of a congruence of null geodesics at any point on $\mathcal{C}$ is negative, and
\item $R_{ab} k^a k^b\geq0$ for any null vector $k^a$. 
\end{enumerate}
The second condition is equivalent, by Einstein's equations, to the null energy condition, which we take to hold (notice that unlike e.g. the strong energy condition, the null energy condition is marginally satisfied by a positive cosmological constant). 

The first point can be established by noting that locally in a small neighbourhood of any point in $\mathcal{C}$, the situation is the one described in the main text: The electric field decays locally to radiation, and immediately afterwards we have a FRWL metric. Using the ansatz \eq{2dmetr} (allowing for angular dependence as well), a null radially affinely parametrized congruence  of ingoing/outgoing null geodesics is given by (see \cite{Ellis:2003mb})
\begin{equation}K_{\pm}^a=\frac{r}{a(\tau)}\left(1,\pm a(\tau),0,0\right).\end{equation}
The corresponding expansions $\theta_{\pm}\equiv \nabla_a K^a_{\pm}$ are
\begin{equation}\theta_{\pm}=\frac{1}{ra}\left(\dot{(r^2)}\pm \partial_x(r^2a)\right).\end{equation}
Exactly on the Nariai branch, the second term vanishes, and the first does as well if there are no charged particles. In the adiabatic approximation, however, $\dot{(r^2)}$ becomes negative at any time $>0$, leading to contraction for both congruences as long as the gradients $\partial_x(r^2a)$ are small enough (i.e. close enough to the Nariai branch). 

Since these arguments work for any point of the initial surface, every null geodesic must be incomplete -- no one is escaping to a hypothetical de Sitter region. The crunch is unavoidable.

It is also interesting to consider why the singularity theorem does not lead to the same conclusion in the quasistatic case. The null energy condition is famously not satisfied by Hawking radiation \cite{Candelas:1980zt}, and perhaps this is the case for the Schwinger radiation as well (we have not checked). It is also not straightforward  to argue that any null congruence of geodesics has negative expansion in the same way as above; up in the Nariai branch, the stress-energy of the produced pairs is localized on the angular $S^2$.

\bibliographystyle{utphys}

\bibliography{dSChargedBHRefsv2}

\end{document}